\documentclass[fleqn,usenatbib]{mnras}

\usepackage{newtxtext,newtxmath}

\usepackage[T1]{fontenc}

\usepackage{graphicx}   
\usepackage{amsmath}    
\usepackage{microtype}
\usepackage{comment}
\usepackage[font=small,skip=5pt]{caption}
\usepackage{xcolor}

\title[Optical Colours of Novae]{Revisiting the Classics: On the Optical Colours of Novae as Standard Crayons}

\author[P. Craig et al.]{Peter Craig,$^{1}$\thanks{E-mail: craigpe1@msu.edu}
Elias Aydi,$^{2,1}$
Laura Chomiuk,$^{1}$
Jay Strader,$^{1}$
Ashley Stone,$^{3}$
Kirill V. Sokolovsky,$^{4}$
\newauthor
Koji Mukai,$^{5,6}$
Adam Kawash,$^{1}$
Joan Guarro Fl\'o,$^{7}$
Christophe Boussin,$^{8}$
St\'ephane Charbonnel,$^{9}$
\newauthor
and Olivier Garde,$^{10}$
\\
$^{1}$Center for Data Intensive and Time Domain Astronomy, Department of Physics and Astronomy, Michigan State University, East Lansing, MI 48824, USA\\
$^{2}$Department of Physics \& Astronomy, Texas Tech University, Box 41051, Lubbock, TX, 79409-1051, USA\\
$^{3}$Department of Physics and Astronomy, West Virginia University, P.O. Box 6315, Morgantown, WV 26506, USA\\
$^{4}$Department of Astronomy, University of Illinois at Urbana-Champaign, 1002 W. Green Street, Urbana, IL 61801, USA\\
$^{5}$Center for Space Science and Technology, University of Maryland Baltimore County, Baltimore, MD 21250, USA\\
$^{6}$CRESST and X-ray Astrophysics Laboratory, NASA/GSFC, Greenbelt MD 20771 USA\\
$^{7}$ Piera Remote Observatory, C/. Balmes 2, 08784 PIERA (Barcelona)\\
$^{8}$ Observatoire de l'Eridan et de la Chevelure de Bérénice, F-02400 Epaux-Bézu, France\\
$^{9}$Durtal Observatory, Durtal, France\\
$^{10}$Observatoire de la Tourbière, 45 Chemin du Lac - 38690 Chabons - France\\
}
\date{Accepted XXX. Received YYY; in original form ZZZ}

\pubyear{2024}

\begin{document}
\label{firstpage}
\pagerange{\pageref{firstpage}--\pageref{lastpage}}
\maketitle


\begin{abstract}
We present a systematic study of the $BVRI$ colours of novae over the course of their eruptions. Where possible, interstellar reddening was measured using the equivalent widths of Diffuse Interstellar Bands (DIBs). Some novae lack spectra with sufficient resolution and signal-to-noise ratios; therefore, we supplement as necessary with 3D and 2D dust maps. Utilising only novae with DIB- or 3D-map-based $E(B-V)$, we find an average intrinsic $(B-V)_0$ colour of novae at $V$-band light curve peak of 0.20 with a standard deviation of 0.31, based on 25 novae. When the light curve has declined by 2 magnitudes ($t_2$), we find an average $(B-V)_0 = -0.03$ with a standard deviation of 0.19. These average colours are consistent with previous findings, although the spreads are larger than previously found due to more accurate reddening estimates. We also examined the intrinsic $(R-I)_0$ and $(V-R)_0$ colours across our sample. These colours behave similarly to $(B-V)_0$, except that the $(V-R)_0$ colour gets redder after peak, likely due to the contributions of emission line flux. We searched for correlations between nova colours and $t_2$, peak $V$-band absolute magnitude, and GeV $\gamma$-ray luminosity, but find no statistically significant correlations. Nova colours can therefore be used as standard ``crayons" to estimate interstellar reddening from photometry alone, with 0.2--0.3 mag uncertainty. We present a novel Bayesian strategy for estimating distances to Galactic novae based on these $E(B-V)$ measurements, independent of assumptions about luminosity, built using 3D dust maps and a stellar mass model of the Milky Way. 

\end{abstract}




\section{Introduction}

A classical nova is a thermonuclear eruption on the surface of a white dwarf, triggered by accretion of non-degenerate material from a companion star \citep{Gallagher1978,BodeEvans,ChomiukReview}. Nova eruptions are known to be individualistic, with many properties varying significantly from one nova to another. Nova light curves are observed to have widely varying decay time-scales, and may either decline smoothly, or include substantial variability, flaring, and/or dust dips \citep{Payne-Gaposchkin1957,BodeEvans,Strope2010}. Spectroscopy of novae implies ejecta velocities ranging over an order of magnitude, from hundreds to thousands of kilometres per second \citep{1895Obs....18..436P,McLaughlin1944,Aydi2020}. Meanwhile, the peak absolute magnitudes of nova eruptions can be described by a normal function with standard deviation of $\sim$1\,mag, with a mean of $\sim -7.2$ \citep{Shafter2017, Schaefer2022, Clark2024}.

The photometric colours of novae may have utility for measuring distances of novae, a generally difficult task. Nova distances have been measured utilizing a wide variety of techniques, such as expansion parallaxes, the Maximum Magnitude versus Rate of Decline (MMRD) relations, and the absolute magnitude at peak (see \cite{Ozdonmez2018,Schaefer2022} for compilations of heterogeneous nova distances and details about these distance measurement techniques). Expansion parallaxes depend on assumptions of spherical symmetry and constant velocity expansions of the nova shells, both of which are often faulty and lead to inaccurate distances (see \citealt{Schaefer2018} for a comparison between \emph{Gaia} parallax distances and expansion parallaxes). The MMRD has been shown to suffer from significant scatter, leading to inaccurate distance measurements \citep{Shara_etal_2017,Clark2024}. Similarly, utilizing the absolute magnitude at peak as a standard candle to assign distances also suffers from large spreads in the peak absolute magnitudes. With the availability of \emph{Gaia} DR3 parallaxes, some novae have high-quality distances available through this channel \citep{Schaefer2018,Schaefer2022}. However, many of the Galactic binary systems that host novae are too faint in quiescence to have parallaxes available in \emph{Gaia} DR3, and lack accurate alternatives \citep{GaiaDR3,GaiaMission}. If nova colours are similar from one nova to the next, observations of, e.g., $(B-V)$ are a promising avenue for measuring interstellar reddening towards novae, enabling extinction measurements to be easily made using broad-band optical light curves. For Galactic novae, this can be combined with 3D dust maps of the Milky Way (MW) disc (e.g. \citealt{Chen2019,Green2019,Marshall2006,Drimmel2003}) to inform us about the distance to a nova.

\cite{Schmidt1957} measured the average intrinsic $(B-V)_0$ colour at nova peak as 0.35 from a sample of M31 novae. Based on a sample of 7 novae and correcting for interstellar reddening,
\cite{VanDenBergh1987} determined that the average peak colour is $(B-V)_0 = 0.23 \pm 0.06$, with a standard deviation of 0.16. The $(B-V)_0$ colours measured at $t_2$, the time when the nova's optical light curve has decayed by 2 magnitudes from peak, have even lower spreads, with a mean of $-0.02 \pm 0.04$ and a standard deviation of 0.12 mag. Using a sample of 24 novae, \cite{Schaefer2022} recently reported an average intrinsic colour of $(B-V)_0 = 0.11 \pm 0.04$ with a dispersion of 0.19 mag at the optical peak. These colours are 0.12 mag bluer than those reported by \cite{VanDenBergh1987}. Unfortunately, many of the $E(B-V)$ values used to determine these extinction-corrected colours are found by assuming a colour for the nova (e.g. \citealt{Hachisu2014, Ozdonmez2016}), so this analysis is prone to circular reasoning.

Colours of novae tend to follow a predictable pattern as the nova evolves. According to the model of \cite{Hachisu2014}, nova colours are primarily determined by photospheric emission near the optical peak, with minimal contributions from emission lines. As the nova evolves, much of the optical emission results from free-free emission in the ejecta outside the now-receding photosphere. During this phase, the colours should stabilize at the free-free spectral energy distribution. At later times, emission features contribute substantial flux to the optical emission, and are expected to differ between novae (due to differences in e.g., composition, density, etc.). As a result, late-time colours are expected to become more varied. See \cite{Williams1991} for a collection of nebular phase spectra for novae that show many significant emission features.

By analysing recent novae that have $E(B-V)$ measurements from non-photometric sources, such as diffuse interstellar bands (DIBs), it is possible to construct a sample of novae with reliable intrinsic colour measurements. This colour analysis can be done not only at the light curve peak, but throughout the light curve as well, allowing a test of the dispersion of nova colours at peak, $t_2$ and beyond. Understanding the colour distribution throughout the eruption can inform how to best utilise optical photometry for measuring reddening under a variety of conditions. 

This is an ideal time to revisit the colour distributions of novae. Recent novae are more likely to be detected early in their evolution, allowing for a larger sample of novae with photometry near their peak optical brightness. This allows for better constraints on both the peak time and $t_2$, and colour measurements at these times. Opportunities for measuring interstellar reddening to novae have also improved significantly. Many novae now have high-resolution spectra available that allow us to measure $E(B-V)$ using DIBs and other spectral features. Since the calibrations of \cite{VanDenBergh1987}, we have also gained access to \emph{Gaia} parallaxes towards some novae \citep{Schaefer2022,Schaefer2018} and 3D maps of dust in the MW \citep{Green2019,Marshall2006,Drimmel2003}, which can be combined to measure interstellar reddening.

Photometry in bands beyond $B$ and $V$ may also be able to provide reasonable estimates of the extinction along the line of sight. $R$ and $I$ band observations are relatively common additions to $B$ and $V$ photometry in recent observations of novae, so colours utilizing $BVRI$ photometry will be the focus of this work. Although SDSS filters have become commonly used in many fields, nova photometry tends to primarily be obtained using the Johnson-Cousins filters. Novae are bright optical transients, and are frequently observed by amateur astronomers utilising this filter system. The addition of the $R$ and $I$ band photometry allows the inclusion of redder bands that are less sensitive to the line of sight extinction, the largest source of uncertainty in colour estimation. 

In \S \ref{sec:sample} of this paper, we describe how we select the sample of novae for our colour analysis. In \S \ref{sec:ebv}, we describe how we estimated the interstellar reddening to these novae, and compare the $E(B-V)$ values estimated from several techniques. In \S \ref{sec:colour}, we detail how our measurements of nova colour are carried out. We present distributions of nova colours at optical peak and $t_2$ in \S \ref{sec:colourdist}, explore if colours correlate with other nova properties in \S \ref{sec:correlation}, and investigate how nova colours change with time in \S \ref{sec:colourev}. Finally, we use our nova colour calibration to estimate $E(B-V)$ and distance (via 3-dimensional dust maps) for a larger sample of novae in \S \ref{sec:distance}, and summarize our findings in \S \ref{sec:concl}. 

\section{Nova Sample and Observations}\label{sec:sample}

To build the sample of novae used in this paper, we begin with a list of 149 Galactic nova eruptions between 2008 and 2022\footnote{\url{https://asd.gsfc.nasa.gov/Koji.Mukai/novae/novae.html}}. To be included in our final sample, a nova must have light curves available in bands appropriate for measuring $(B-V)$, $(R-I)$, and/or $(V-R)$. Only novae for which the time of peak can be identified are included. Our measurement of the peak time for each nova is made using the $V$-band light curves, with the uncertainty in the peak time derived from the cadence around peak. Following Craig et al.\ 2025 (in preparation), we exclude novae with uncertainty in the time of peak greater than eight days or $\frac{1}{3}t_2$.

If we do not have photometry capturing at least part of the rise to peak, then the nova will also be excluded due to uncertainty in the peak time. Our derived peak times, and their uncertainties, are listed in Table \ref{tab:allnovae}. Despite allowing for peak times uncertainties of up to eight days, the majority of our sources have peak time uncertainties of less than two days. If the peak is detected, $t_2$ is measured by linearly interpolating along the $V$-band light curve, and finding the last time that the nova falls 2 magnitudes from the peak, matching the methods of Craig et al.\ 2025 (in preparation). This is done across all the available $V$ band photometry, and does not utilise data in other bands. If nova colours were to evolve between peak and $t_2$, then the derived $t_2$ might be different in different bands, so we use $V$ band everywhere for consistency. We note that \cite{VanDenBergh1987} defined $t_2$ as the first time that the light curve falls 2 magnitudes below peak, but we instead use the last time that the nova falls below this magnitude (in keeping with e.g., \citealt{Strope2010}). The resulting sample includes 61 novae that are used for at least one colour measurement.

\begin{figure}
 \includegraphics[width=\columnwidth]{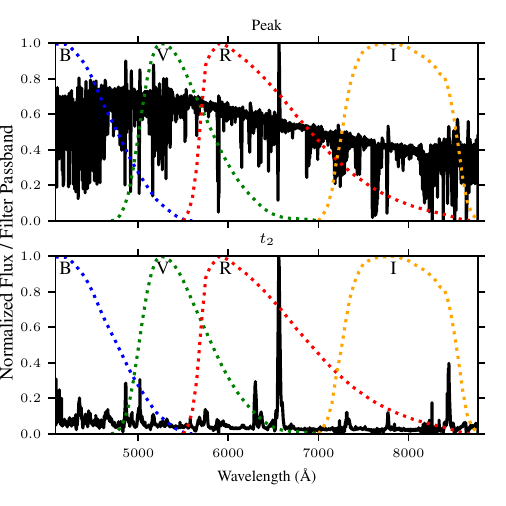}
     \caption{Normalized spectra with filter curves overlaid for the nova FM Cir. Both spectra are SMARTS Chiron spectra, the first (top panel) taken near the nova peak on 2018-01-28 and the second (bottom panel) taken near $t_2$ for this nova on 2018-06-15.}
 \label{fig:samplespec}
\end{figure}

All of our photometry is obtained through either the American Association of Variable Star Observers (AAVSO; \citealt{Kloppenborg23}) international database, or with the Small and Medium Aperture Telescope System (SMARTS). The SMARTS data were primarily obtained using ANDICAM on the 1.3\,m telescope, and were publicly released as part of the Stony Brook/SMARTS Atlas of (mostly) Southern Novae \citep{Walter2012}. 
For this work, we focus on $BVRI$ photometry, as these four bands are the most commonly available (we investigate the optical--infrared colours of novae, e.g., $(V-K)$, in Appendices A and B of \cite{Chong2025}. SMARTS photometry typically includes all four bands, while AAVSO photometry may not, and is more likely to have $B$ and $V$ bands. $V$ band in particular tends to be the typical band for nova photometry, as it requires short exposures compared to $B$ band and avoids H$\alpha$, which is useful for tracking the continuum emission. See Figure \ref{fig:samplespec} for sample spectra for the nova FM Cir at peak and at $t_2$, with the $BVRI$ filter curves overlaid.

Spectra for the novae in this sample are primarily obtained through the Astronomical Ring for Access to Spectroscopy (ARAS; \citealt{Teyssier_2019}\footnote{\url{https://aras-database.github.io/database/index.html}}). The data used for our work consists primarily of medium-resolution spectroscopy ($R\gtrsim 9,000$), typically covering the wavelength range of $\sim4,000\text{ \AA}$ to $\sim7,500\text{ \AA}$. Some spectra limited to within $\sim 200\text{ \AA}$ of H$\alpha$ are included as well, as they often cover the DIB at $6613\text{ \AA}$. We also utilise SMARTS CHIRON spectra \citep{Walter2012}, which 
are echelle spectra with resolution of $R\approx27,000$ that cover a broad wavelength range from $4080\text{ \AA}$ to $8910\text{ \AA}$. A small number of published spectra obtained using the High Resolution Spectrograph (HRS; \citealt{Barnes_etal_2008,Bramall_etal_2010,Bramall_etal_2012,Crause_etal_2014}) mounted on the 11-m Southern African Large Telescope (SALT; \citealt{Buckley_etal_2006,Odonoghue_etal_2006}) are also included in the sample.

\section{\texorpdfstring{$E(B-V)$}{E(B-V)} Measurements}\label{sec:ebv}

Colour measurements for this sample depend heavily on estimating the colour excess from interstellar reddening. This can vary significantly in strength from one nova to another, and is especially relevant for novae close to the Galactic plane. This section describes the methodology that we use for measuring $E(B-V)$ towards the novae in our sample. In order to extend this to colours other than $(B-V)$, we use the extinction law presented in \cite{Wang2019} to estimate the appropriate colour excess. For the colours measured here, this extinction law gives $E(R-I) = 0.75\, E(B-V)$ and $E(V-R) = 0.66\, E(B-V)$.

\subsection{$E(B-V)$ from Diffuse Interstellar Bands}\label{sec:dib}

For novae where spectra are available with resolutions $R \gtrsim 9,000$, we use measurements of the equivalent widths (EWs) of DIBs to measure the reddening due to interstellar material, $E(B-V)$. DIBs have been detected in stellar spectra since as early as 1919 \citep{Heger1922}. DIBs are interstellar absorption features produced by unknown (possibly carbon-based) carriers \citep{Herbig1995,Snow2006} along the line of sight. The one exception is the fullerene ion C$_{60}^{+}$, which has been associated with several DIBs \citep{Campbell2015,Walker2015}. DIB carriers have been found to be correlated with both the gas and dust, and therefore reddening. Some DIB EWs (such as the DIB at $5780.5$ \AA{}) show stronger correlations with the column densities of H than with $E(B-V)$ \citep{Friedman2011}. Calibration accuracy depends on the consistency of the ratio between the unknown DIB carriers and the ISM dust. For a review of DIBs, see \cite{Herbig1995,Snow2006}.

DIB EWs are correlated with $E(B-V)$ along the line of sight \citep{Merrill1938,Friedman2011,Vos2011,Lan2015}. We follow the relations provided in \cite{Friedman2011} between $E(B-V)$ and the EW of eight DIBs between $5485\text{ \AA}$ and $6615 \text{ \AA}$. These relationships are established by measuring the EWs for each of these eight DIBs in the spectra of 133 stars. Each star has an independent reddening measurement derived from intrinsic stellar colours, and the calibrations are derived as the best fit lines between the DIB EW and the independent $E(B-V)$. In our nova sample, we measure the EWs for all eight DIBs when possible, although typically only a subset of these are detected, inside the covered wavelength range, and free from significant line blending. The most commonly included DIBs are those with central wavelengths at 5780.5 \AA{}, 5797.1 \AA{} and 6613.6 \AA{}. We note that the DIBs covered here are rarely seen to have multiple velocity components in the available spectra, likely because the spectra lack sufficient resolution to detect these small velocity effects. Some spectra do show multiple velocity components in the \ion{Na}{I} lines, but they are not evident in the broad DIB features.

EW measurements are made in \textsc{Python} by directly integrating over the absorption features. The first step in our method is to determine a region that belongs to the continuum around the line. This begins with all data within 20 \AA{} of the central wavelength ($\lambda_c$) of the DIB, excluding data within 2 \AA{} of $\lambda_c$. In the case of the DIBs at 5780.5 \AA{} and 5797.1 \AA{}, these features are within 20 \AA{} of each other, so the continuum data for one of these lines will include the other. To avoid this, all data within 2 \AA{} of a known DIB are removed from the continuum. Each region is also inspected manually, to ensure that there are not other spectral lines present in the identified continuum region (or worse, blended with the DIBs leading to overestimated EW measurements). Spectral lines in the continuum fit can be removed manually, and DIBs suffering from line blending are not included in our analysis. With this continuum selection, a continuum model is generated by fitting a 6th order Legendre polynomial to the continuum.

To determine the line edges used for integration, we take the first point on each side of $\lambda_c$ where the continuum model and the spectrum cross. Our direct integration is intended to be similar to the \textsc{IRAF} splot command, so we integrate under a linear continuum fit. The end points for our linear continuum are 0.7 \AA{} beyond the derived line edges, and the flux for each endpoint is the average of all continuum data within 1.4 \AA{} of the line edge (excluding the region between the two line edges). This method has been used before for EW measurements, and reduces the sensitivity of the EW to precise placement of the line edges compared to directly integrating between the derived edges \citep{Vos2011}. Our EW values are then computed by integrating under this linear continuum.

In a few cases, the continuum-fit derived line edges lead to unreasonable models for this local continuum fit. Typically, this occurs due to a low-quality continuum model, which leads to line edges being assigned far away ($\gtrsim 0.2$ \AA{}) from the actual DIB edges. The most common culprit for this issue is the DIB at 6613 \AA{}, due to H$\alpha$ emission around this feature, although occasionally it happens with other DIBs as well. Each line is checked manually to ensure that the derived fit is reasonable, and that the selected line edges are close to the ones that would be selected manually (as would be done in an analysis with the \textsc{IRAF} splot command). In cases where the change in the line edge position would make a significant impact on the estimated EW, new line edges are selected manually. The automated case provides the benefit of being consistent across many sources and is a repeatable measurement, but occasionally requires human intervention.

Uncertainties on the resulting EW measurements are determined following the analysis in \citet{Vollmann2006}, making use of equation 7 from that paper, reproduced here:
\begin{equation}\label{eqn:uncertainty}
\sigma(W_\lambda) = \sqrt{1 + \frac{\overline{F_c}}{\overline{F}} } \frac{\Delta \lambda - W_\lambda}{S/N}
\end{equation} 
Here $\overline{F}$ is the average flux in the line, $\overline{F_c}$ is the average continuum flux, $\Delta \lambda$ is the width of the integration, $W_\lambda$ is the measured EW, and S/N is the signal-to-noise ratio of the continuum. Our S/N is computed by taking the average flux in the continuum around the line (the same continuum data used for continuum fitting), and dividing by the RMS of this data after subtracting the continuum model. This method gives reasonable agreement with \textsc{IRAF} splot signal-to-noise measurements.

We measure the EW for each clearly detected DIB in all available spectra for the nova. For each individual DIB, we then compute a weighted mean EW across all measurements, and propagate the measurement uncertainties using standard error propagation. This procedure yields a measurement of the EW, with an uncertainty, for each DIB available in the data. To check for variability in DIB EWs between spectra, we have compared the EW measurements for each DIB across all spectra for each nova. The EW is consistent with the final measurement of the EW to within the uncertainties $94\%$ of the time.

Next, we calculate $E(B-V)$ for each DIB using the calibration of \cite{Friedman2011}. Uncertainties caused by the scatter in the EW vs. $E(B-V)$ relations are accounted for, and determined based on the data in Table 1 of \cite{Friedman2011}. For each DIB, measurements are sorted by EW, and then split into bins each containing 30 measurements. The final bin may contain more than 30 measurements, as it includes any larger EW measurements that will not form a full bin on their own. The scatter in each bin is quantified as the standard deviation of the $E(B-V)$ values predicted by the EWs. In our DIB measurements, the final uncertainty is the scatter appropriate for the measured EW combined with the uncertainty from the EW measurement in quadrature. The scatter in the DIB relationships is typically $\sim 0.3$ mag, while the uncertainty on $E(B-V)$ from the EW uncertainty is $\sim 0.1$ mag.

Finally, all the individual $E(B-V)$ measurements are then combined using a weighted average, with weights set as $1/\sigma^2$, and uncertainties are propagated with standard uncertainty propagation. 
The \textsc{Python} routine used for these DIB measurements is publicly available\footnote{\url{https://github.com/AdkPete/reddening-dibs}}. All of our derived E(B-V) measurements, with uncertainties where available, can be found in Table \ref{tab:allnovae}. This table includes all the novae used in our colour sample, and includes their intrinsic colours at peak and $t_2$.

For a few sources, there are measurements of the EWs for various DIBs available in the literature, or there are DIB-based measurements of $E(B-V)$ available. We adopt these measurements for a small number of sources in our sample, where the publicly available spectra have relatively low resolution (or lack DIB detections), in which case they will likely provide inaccurate measurements compared to the literature values. When DIB EWs are available, we use the same technique to combine the DIBs into an $E(B-V)$ value that we use with our own DIB measurements. T Pyx has EW measurements available from high resolution spectra from the Nordic Optical Telescope (NOT) \cite{Shore2011}. V1324 Sco has EWs available from \cite{Finzell2015}, measured using spectra from the Very Large Telescope (VLT) and the Magellan telescope. V407 Lup also has EW measurements using spectra from the UVES spectrograph mounted on the VLT \citep{Izzo2018}. Finally, we adopt the $E(B-V)$ value for V5589 Sgr from \cite{Weston2016}, obtained using the Tillinghast Reflector Echelle Spectrograph mounted on the Fred L. Whipple Observatory. Many other novae have spectroscopically derived $E(B-V)$ values available in the literature using a variety of techniques. However, EWs are rarely published, and we do not typically include these estimates, favouring instead to homogeneously apply our methods with publicly available spectra to be as consistent as possible.

\subsection{$E(B-V)$ from \ion{Na}{i} D Absorption Lines}\label{sec:naid}

In addition to measurements of DIB EWs, the EW of the \ion{Na}{i} D absorption features (both lines of the doublet) are measurable from the spectra following the procedure in section \ref{sec:dib}. \ion{Na}{i} D line EWs have also been shown to be correlated with $E(B-V)$, and we use the calibrations of \cite{Poznanski2012}. When possible, we make this measurement of $E(B-V)$ as well to compare against the DIB values. For sources with large $E(B-V)$, the \ion{Na}{i} D lines will be saturated, and reliable $E(B-V)$ measurements from the spectra are only available from DIBs. This is typically problematic for sources with $E(B-V)$ $\gtrsim 1.5$, although depending on the properties of the absorbing clouds along the line of sight, this may occur at lower $E(B-V)$ \citep{Poznanski2012}.

\begin{figure}
 \includegraphics[width=\columnwidth]{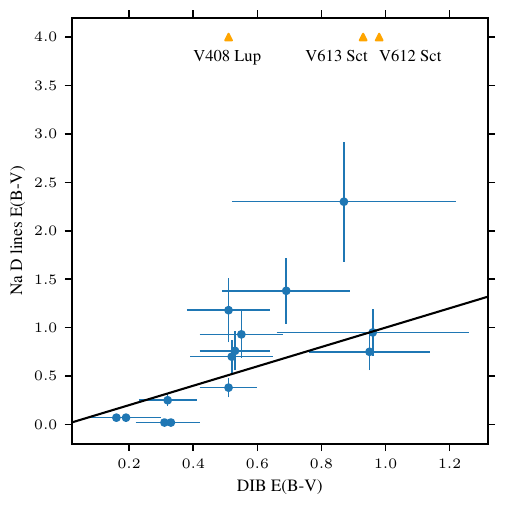}
 \caption{$E(B-V)$ values predicted by DIB measurements plotted against $E(B-V)$ estimated with the \ion{Na}{i} D doublet. The solid line indicates where the two values are equal. The orange triangles represent sources with greatly overestimated $E(B-V)$ from the \ion{Na}{i} measurements. V408 Lup has an \ion{Na}{i} D $E(B-V)$ of 9.25, V613 Sct's estimate is at 608, and V612 Sct is at 17.4. Generally, we find approximate agreement between the two methods, with some sources having overestimated $E(B-V)$ from the \ion{Na}{i} D doublet. V613 Sct at $E(B-V) = 608$ suffers from line saturation, but the lower estimates do not appear to be saturated, possibly a result of insufficient spectral resolution.}
 \label{fig:dib_v_Na}
\end{figure}

With the compiled set of measurements available from both DIBs and the \ion{Na}{i} D lines, we can compare the $E(B-V)$ values predicted by the two methods, shown in Figure~\ref{fig:dib_v_Na}. The two methods typically agree, but some sources have much larger values for $E(B-V)$ predicted from the \ion{Na}{i} D lines. In extreme cases this leads to unrealistically large estimates of $E(B-V)$, as in the case of V613 Sct where $E(B-V)$ is estimated to be 608\,mag (likely a result of saturation). Single-component \ion{Na}{i} D profiles are affected by saturation for $E(B-V) \ga 0.4$, flattening the EW--$E(B-V)$ relation at higher extinctions \citep{Munari1997}. If the \ion{Na}{i} D absorption is composed of multiple components, it will take a larger absorbing column to reach saturation, and the EW will be higher than expected from the \cite{Poznanski2012} relation---in turn yielding a higher $E(B-V)$ estimate.  Multiple absorption components may arise from interstellar clouds at different distances along the line of sight, which in general will have different line of sight velocities, producing multiple absorption components.

For these reasons, \citet{Poznanski2011} point out that \ion{Na}{i} D EW is not a good proxy for extinction if measured from low-resolution spectra (which will blend together multiple components and limit tests for saturation).
While the spectra we consider here are all of higher resolution than those considered by \citet{Poznanski2011}, the three novae with very discrepant $E(B-V)$ measurements were all observed in medium resolution ($R \approx 9,000 - 13,000$) spectra. At the higher end of this resolution range, there is evidence of line blending from multiple components in the spectra for V408 Lup.

For these reasons, we do not include the \ion{Na}{i} D line measurements in our final values for $E(B-V)$, and generally prefer DIB-derived reddening values over the \ion{Na}{i} D measurements.

\subsection{$E(B-V)$ from Dust Maps} \label{sec:dustmaps}
When suitable spectra are not available, $E(B-V)$ is instead estimated from dust maps. If an accurate parallax measurement from \emph{Gaia} DR3 \citet{GaiaMission,GaiaDR3} is available, then we use 3D dust maps in combination with distance constraints. Distances are derived from \emph{Gaia} parallaxes, when the nova has a sufficiently accurate \emph{Gaia} DR3 parallax, defined as systems where the parallax error is less than one third of the parallax. The parallax distances (and uncertainties) are derived using the method of \cite{Bailer-Jones2018}, with the length scale prior set following \cite{Schaefer2018}. All dust map $E(B-V)$ values are found using the \textsc{mwdust} python package \citep{Bovy2016}. We use the \verb|Combined19| map from this package, which is a combination of the 3D dust maps from \cite{Green2019}, \cite{Marshall2006} and \cite{Drimmel2003}. The 3D dust maps are only used for five novae in our sample of 61, as most cases with accurate parallaxes also have spectra suitable for DIB measurements. See Table \ref{tab:allnovae} for more specific information, including the $E(B-V)$ value for each nova in the sample, the method used to measure $E(B-V)$, and parallax distance estimates to the novae where available (the ``distance" column is estimated from interstellar reddening as described in \S \ref{sec:distance}).

Typical uncertainties on the $E(B-V)$ from 3D dust map have been estimated to be 0.15 magnitudes across our sample, based on the comparison between 3D dust map $E(B-V)$ and DIB measurements. This is a bit larger than the median uncertainties in \cite{Green2019} of 0.086 mag, possibly including some contributions from small scale variations in the extinctions as well as distance uncertainties. To account for the parallax uncertainties, this 0.15 magnitude uncertainty is combined in quadrature with the uncertainties from the distance, calculated based on the 3D dust map $E(B-V)$ at one sigma below the distance estimate and one sigma above this distance. For some of our 3D map novae this works out to be 0, as the distance minus one sigma remains beyond most or all of the Galactic extinction, in which case the final uncertainty is 0.15 mag.

\begin{figure}
 \includegraphics[]{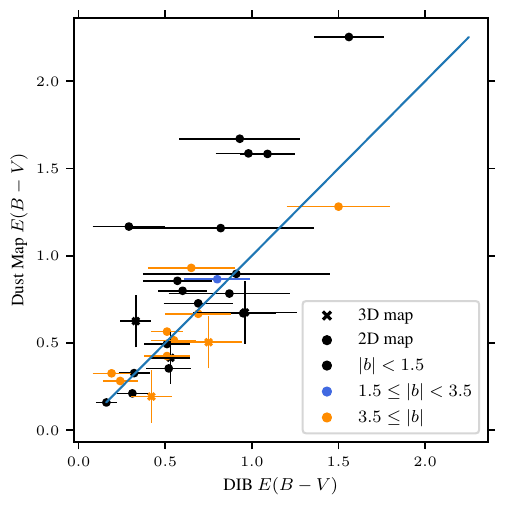}
 \caption{$E(B-V)$ values predicted by DIB measurements plotted against those estimated with dust maps. The points marked as an x use 3D dust maps, while the remaining sources are based on integrated dust maps. Along the solid line, the two values are equal. The black points show novae within 1.5 degree of the plane of the MW, orange points are between 1.5 and 3.5 degrees and blue points are farther than 3.5 degrees.}
 \label{fig:dib_v_Map}
\end{figure}

When DIB or 3D dust map measurements are not available, 2D dust map $E(B-V)$ values are adopted, effectively providing an upper limit on the reddening for 32 novae (see Table \ref{tab:allnovae}). 2D map values are obtained using the map from \cite{Schlafly2011}. In order to investigate how large the overestimation of $E(B-V)$ values from these 2D maps is, we compare our spectroscopically derived DIB measurements with the 2D map values in Figure~\ref{fig:dib_v_Map}. For most of the novae in our sample, we find that the 2D dust maps provide reasonable values for $E(B-V)$ when compared to DIB measurements. We attribute this to many novae lying beyond the majority of the Galactic dust, especially at higher Galactic latitudes. Our main results are derived without these sources, to avoid the potential systematic overestimation in $E(B-V)$ associated with the 2D maps.
In Figure~\ref{fig:dib_v_Map}, it is apparent that most novae have decent agreement between $E(B-V)$ values, within 0.2 magnitudes for the majority of the sample. The most discrepant novae (more than 0.4 magnitudes) are at Galactic latitudes of less than 1.5 degrees, where it is not surprising that significant dust is found at distances beyond the nova.

\section{Colour Measurements} \label{sec:colour}

To measure the colour of a nova at a specific time, we utilise all photometry available within 1.5 days of that time. We select sets of photometry from the two bands that are reported to AAVSO by the same observer (or are all SMARTS measurements) and are within 1.2 hours of each other; usually the measurements in different filters are only separated by a few minutes. Each photometric measurement is only used once, and is paired with the closest available point in the other band. Each pair produces a colour measurement, and these measurements are then combined using a weighted average, with weights determined as $1/\Delta t$, where $\Delta t$ is the time difference between the measurement and the desired time in days. If $\Delta t < 1\ \mathrm{hour}$, we calculate the weight using $\Delta t = 1\ \mathrm{hour}$. This is intended to avoid massive weights being assigned to photometry very close to the optical peak.

For example, if we want to measure the colour at peak for a given nova, we begin with all of the photometry within 1.5 days of peak. Starting with the measurements closest to the peak in one band, we measure a colour using the closest point in the other band. If there is not another point in the other band within 1.2 hours of this point, we move on. Otherwise, these two points are removed, and we continue this until we run out of data in one of the bands. These measurements are then combined through the weighted averaging to produce a single colour measurement.

The bulk of the uncertainty on the colour derives from the reddening uncertainty, which is typically between 0.1 and 0.3 magnitudes in the silver sample. The other significant source of uncertainty is from systematic offsets between different observers, both between individual AAVSO observers and between AAVSO and SMARTS. To characterize this uncertainty, we have compared the photometry reported by different observers, and find typical offsets on the order of 0.1 magnitudes. This is generally a bit larger than the quoted uncertainties, and we adopt an uncertainty component of 0.125 on our colours to account for this, which is then combined in quadrature with the reddening uncertainties to calculate the uncertainty on the colour. The results of this paper do not vary significantly if we restrict our analysis to the photometry from a single observer (removing any systematic offsets). This does lead to smaller sample sizes due to the more limited dataset, but the distributions shown in section \ref{sec:colourdist} are roughly the same when generated using only SMARTS data (or data from just one or two AAVSO observers).

\section{Results \& Discussion} \label{sec:results}

\subsection{Distribution of Nova Colours at Peak and $t_2$}\label{sec:colourdist}

\begin{figure}
 \includegraphics[width=\columnwidth]{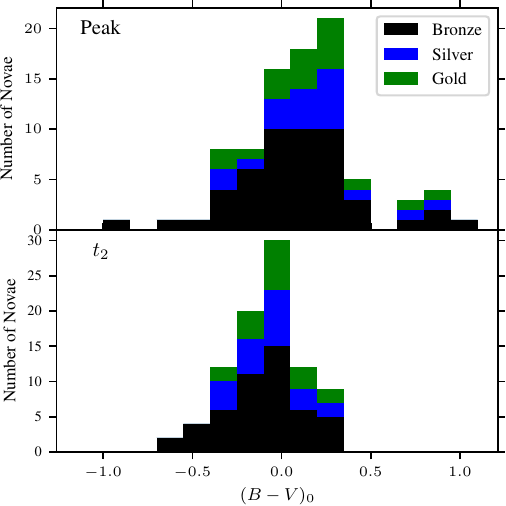}
 \caption{Histograms here display the distributions of $(B-V)_0$ colours for our three samples. The top panel displays the colour at peak, while the bottom panel shows the colour at $t_2$. The green bars show the novae belonging to the gold sample, the blue bars for the silver sample and black bars for the bronze sample. The full distribution shown is then the full nova sample.}
 \label{fig:(B-V)hist}
\end{figure}

We have measured three different colours for the novae in our sample at peak, $t_2$ and throughout their light curves. In this section, we analyse the distributions of these colours at peak and $t_2$. Table \ref{tab:colourstats} contains the mean, median, and standard deviation for each of these distributions. We split our statistics into three separate nova samples by quality of the data: gold, silver, and bronze. The statistics presented throughout the paper are from the silver sample, which provides our fiducial results. The sample used for our figures varies, and is noted in the figure captions.

The gold sample is defined as the set of novae with $E(B-V)$ measurements from DIBs, reliable $t_2$ measurements, and photometry that is available at both peak and $t_2$. In this case, we can compare statistics across both peak and $t_2$ using an identical sample, and only using our best $E(B-V)$ measurements. The weakness of this sample is the limited available sample size. The size of each sample, and the statistics describing the colour distributions for each sample, may be found in Table \ref{tab:colourstats}. The silver sample removes the requirement for photometric availability both at peak and at $t_2$, and includes novae with $E(B-V)$ from 3D dust maps. This means that we are using different samples of novae at peak and at $t_2$, while still using reliable $E(B-V)$ measurements. The bronze sample includes all sources with suitable photometry, and allows 2D dust map estimates for $E(B-V)$.

In the bronze sample, there are typically a few outliers with colours that differ significantly from most novae in the sample, caused by overestimated $E(B-V)$ values derived from 2D dust maps. To reject these sources, we remove any novae more than three standard deviations from the mean, then recalculate the mean and standard deviation. This procedure is repeated until there are no outliers remaining. All figures and statistics generated using the bronze sample have this cut applied. We do not apply any outlier rejection to the silver or gold samples, since they do not include integrated dust map reddenings.

The distribution of $(B-V)_0$ colours at peak and at $t_2$ are shown in Figure~\ref{fig:(B-V)hist}, including all of the novae in our bronze sample except for those removed during outlier rejection. At peak, the silver sample has a mean $(B-V)_0$ of 0.20 with a standard deviation of 0.31, and includes a total of 25 novae. The average colour in the bronze sample is slightly bluer at 0.08, likely due to a small systematic overestimation of $E(B-V)$ from the 2D dust maps. The sample plotted in Figure~\ref{fig:(B-V)hist} includes all novae in our sample, split up based on which subsample they belong to. The $(B-V)_0$ colours at $t_2$ show a smaller spread compared to those at the peak. At $t_2$, we find an average $(B-V)_0$ of $-0.03$ with a standard deviation of 0.19. The sample of novae at $t_2$ is restricted by the requirement of a reliable estimate of $t_2$, though most novae with parallaxes or high resolution spectra for DIB measurements have well sampled light curves, and therefore reliable $t_2$ values.

\begin{figure}
 \includegraphics[width=\columnwidth]{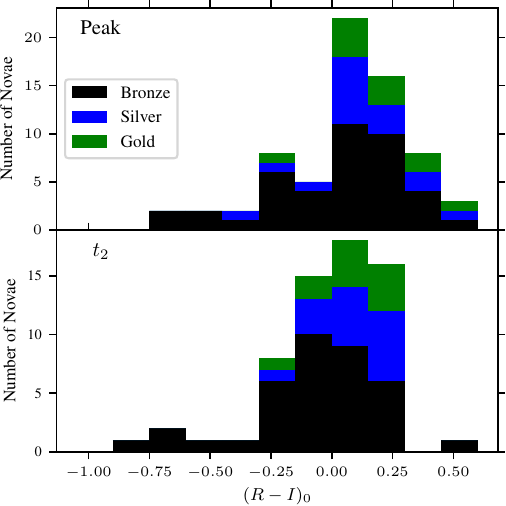}
 \caption{Histograms here display the distributions of $(R-I)_0$ colours for our three samples. The top panel displays the colour at peak, while the bottom panel shows the colour at $t_2$. The green bars show the novae belonging to the gold sample, the blue bars for the silver sample and black bars for the bronze sample. The full distribution shown is then the full nova sample.}
 \label{fig:(R-I)hist}
\end{figure}

\begin{figure}
 \includegraphics[width=\columnwidth]{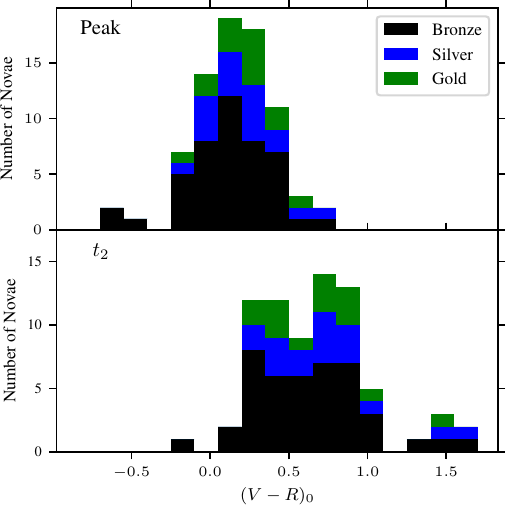}
 \caption{Histograms here display the distributions of $(V-R)_0$ colours for our three samples. The top panel displays the colour at peak, while the bottom panel shows the colour at $t_2$. The green bars show the novae belonging to the gold sample, the blue bars for the silver sample and black bars for the bronze sample. The full distribution shown is then the full nova sample.}
 \label{fig:(V-R)hist}
\end{figure}

Distributions of intrinsic $(R-I)_0$ colours for our sample are shown in Figure~\ref{fig:(R-I)hist}, both at peak and at $t_2$. Availability of $R$ and $I$ band photometry is somewhat limited compared to the $B$ and $V$ bands, so the sample size at peak is reduced compared to $(B-V)$. Based on the silver sample, the spread of the $(R-I)_0$ data at peak is 0.19, which is lower than the spread in $(B-V)_0$, and with an average $(R-I)_0$ of 0.13. At $t_2$, the standard deviation in the $(R-I)_0$ sample is 0.27, with an average  $(R-I)_0$ of 0.10, based on a total of 20 novae. The final colour that we check is $(V-R)$, with the distributions for our sample at peak and at $t_2$ shown in Figure~\ref{fig:(V-R)hist}. At peak, the average $(V-R)_0$ for the silver sample is 0.20, with a standard deviation of 0.23 and a sample size of 23 novae. At $t_2$, the mean becomes 0.68 with a standard deviation of 0.40, and a sample size of 22 novae.

For investigators looking for a ``typical" nova colour or characterization of colours in the general nova population, \textbf{we recommend using values derived from the silver samples}, tabulated in Table \ref{tab:colourstats}. While the gold samples are ideal for some comparisons between the colours at peak and $t_2$, the silver sample provides the best statistics on colours at particular times in the eruption (i.e., peak, $t_2$) thanks to the larger sample sizes.

One potential concern for this analysis is the accuracy of the derivation of peak times for these samples. Given that our primary results are derived from the silver sample, we reconsider these statistics restricting the analysis to the subset of silver sample novae that have small peak uncertainties. Here we are taking the larger of the asymmetric uncertainties for all novae with asymmetric uncertainties on the peak time. Of the 25 silver sample novae with $(B-V)_0$ photometry at peak, 24 of them have peak time uncertainties of less than 2 days, and 20 have uncertainties of less than 1 day. For these 20 novae, we recover an average $(B-V)_0$ at peak of 0.20 with standard deviation of 0.33. Similarly, for the novae with peak time uncertainties of less than 1 day, we recover an average $(R-I)_0$ = 0.13, with a standard deviation of 0.21 based on 16 novae. In $(V-R)_0$, the average becomes 0.19 with standard deviation 0.24 and a sample size of 19.

At $t_2$, this same nova sample with peak uncertainties less than 1 day yields very similar results to our full sample. We recover an average $(B-V)_0$ at $t_2$ of -0.05 with a standard deviation of 0.19, based on 21 novae. For $(R-I)_0$ at $t_2$, we see an average of 0.09 with a standard deviation of 0.21, with a sample size of 16 novae. In $(V-R)_0$, this sample has an average of 0.70 with a standard deviation of 0.43, based on 18 novae. These stats are in excellent agreement with the full sample, leading us to conclude that the relevant statistics in our full sample are not heavily impacted by the uncertainty on the peak time.

In the model of \cite{Hachisu2014}, the colour at $t_2$ is thought to be driven primarily from free-free emission of the ejecta outside the photosphere. They do not consider colours after $t_3$ (when the nova has declined by three magnitudes), due to the contributions of the emission lines to the optical flux. Based on this model, the colours at $t_2$ should be consistent from one nova to another. This is generally supported by our data for the $(B-V)_0$ colours, with somewhat larger variations in colour appearing at peak compared to $t_2$. Compared to previous results, we get comparable values for colours at both peak and $t_2$, although with a larger standard deviation than other authors find. Our $(B-V)_0$ colour at peak is a bit redder than that presented in \citet{Schaefer2022}, which is not necessarily unexpected given the amount of variability in our samples and the available number of novae. Our average $(B-V)_0$ colours at both peak and $t_2$ agree with those measured in \citet{VanDenBergh1987}. More notably, we recover a larger standard deviation than both \citet{Schaefer2022} and \citet{VanDenBergh1987} at both times.

One possible contributor to the colour distribution at $t_2$ is the presence of dust formation in novae. Some novae show large dust dips in their optical light curves due to extinction from newly formed dust, and in some cases the sudden drop in optical flux can be what determines $t_2$ (see \cite{Chong2025} submitted for details). In our silver sample, there are nine novae that exhibit significant dust dips that may impact the timing of $t_2$. The colours of the dusty novae at $t_2$, of which 7 have suitable photometry, show a standard deviation of 0.22 with a mean colour of 0.08. This is similar to the full sample, though a bit redder, and the non-dusty novae also show statistics similar to the full sample (although with a somewhat bluer mean of -0.09). The sample sizes here are quite limited however, and while it appears that dusty novae are a bit redder than non-dusty novae, we cannot detect any significant differences in the $t_2$ colours between the dusty and non-dusty samples. There are not any significant differences between these samples for $(R-I)_0$ or $(V-R)_0$ either.

\begin{table}
\caption{Statistics for our analysed intrinsic colours across the gold, silver and bronze samples. The gold sample consists of novae that have DIB measurements for $E(B-V)$, and have photometry available at both peak and $t_2$. The silver sample is restricted to the novae that have DIB or 3D dust map based measurements for $E(B-V)$, but includes sources that are missing photometry at either peak or $t_2$. The bronze sample includes all of our novae with photometry at peak or $t_2$. \textbf{We recommend that the nova community use values estimated from the silver samples, which includes the highest quality $E(B-V)$ measurements over the largest sample size.}}\label{tab:colourstats}
\begin{tabular}{|c|c|c|c|c|c|}
\hline
Sample & N & Mean & +/- Mean & Median & $\sigma$ \\ 
\hline
 $(B-V)_0$ Peak Bronze & 50 & 0.08 & 0.05 & 0.10 & 0.36 \\ 
 $\mathbf{(B-V)_0}$ \bf{Peak Silver} & \bf{25} & \bf{0.20} & \bf{0.06} & \bf{0.16} & \bf{0.31} \\ 
 $(B-V)_0$ Peak Gold & 18 & 0.18 & 0.08 & 0.16 & 0.31 \\ 
 $(B-V)_0\ t_2$ Bronze & 49 & $-0.11$ & 0.03 & $-0.08$ & 0.22 \\ 
 $\mathbf{(B-V)_0\ t_2}$ \bf{Silver} & \bf{27} & $\mathbf{-0.03}$ & \bf{0.04} & $\mathbf{-0.06}$ & \bf{0.19} \\ 
 $(B-V)_0\ t_2$ Gold & 18 & $-0.03$ & 0.04 & $-0.02$ & 0.17 \\ 
 \hline
 $(R-I)_0$ Peak Bronze & 41 & 0.02 & 0.04 & 0.04 & 0.27 \\ 
 $\mathbf{(R-I)_0}$ \bf{Peak Silver} & \bf{20} & \bf{0.13} & \bf{0.04} & \bf{0.14} & \bf{0.19} \\ 
 $(R-I)$ Peak Gold & 11 & 0.17 & 0.06 & 0.16 & 0.18 \\ 
 $(R-I)_0\ t_2$ Bronze & 37 & $-0.08$ & 0.05 & $-0.02$ & 0.28 \\ 
 $\mathbf{(R-I)\ t_2}$ \bf{Silver} & \bf{20} & \bf{0.1} & \bf{0.06} & \bf{0.06} & \bf{0.27} \\ 
 $(R-I)\ t_2$ Gold & 11 & 0.06 & 0.05 & 0.08 & 0.16 \\ 
 \hline
 $(V-R)_0$ Peak Bronze & 45 & 0.11 & 0.04 & 0.14 & 0.28 \\ 
 $\mathbf{(V-R)_0}$ \bf{Peak Silver} & \bf{23} & \bf{0.20} & \bf{0.05} & \bf{0.17} & \bf{0.23} \\ 
 $(V-R)_0$ Peak Gold & 14 & 0.19 & 0.06 & 0.21 & 0.20 \\ 
 $(V-R)_0\ t_2$ Bronze & 43 & 0.63 & 0.06 & 0.60 & 0.36 \\ 
 $\mathbf{(V-R)_0\ t_2}$ \bf{Silver} & \bf{22} & \bf{0.68} & \bf{0.09} & \bf{0.69} & \bf{0.40} \\ 
 $(V-R)_0\ t_2$ Gold & 14 & 0.71 & 0.08 & 0.70 & 0.31 \\ 
 \hline
\end{tabular}
\end{table}

\begin{figure}
 \includegraphics[width=\columnwidth]{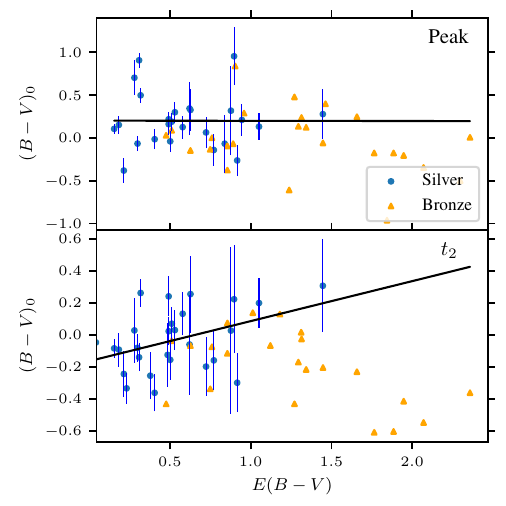}
 \caption{ The top panel displays the peak $(B-V)_0$ plotted against the estimated $E(B-V)$ values. The bottom panel is similar, but displays the colours at $t_2$. The black line displays a linear best fit line made to the silver sample in each case. Orange triangles in these figures use $E(B-V)$ measurements made using dust map values, and the blue points with uncertainties are based on DIB measurements. All sources combined represent the bronze sample, and the blue points are the silver sample. The fit at peak to the silver sample has a p-value of 0.46 and a Pearson correlation coefficient of -0.19. At $t_2$, the p-value for the silver sample is 0.04, with a Pearson correlation coefficient of 0.40. This correlation is largely driven by the nova in the silver sample with the largest reddening, and is likely due to an underestimate of the reddening for this source. Without this source, the correlation disappears.}
 \label{fig:bv_vs_ebv}
\end{figure}

To test for systematics in our $E(B-V)$ measurements, we check for correlations between the derived intrinsic colours and the estimated $E(B-V)$. In Figure~\ref{fig:bv_vs_ebv} we show a scatter plot of the $(B-V)_0$ colours plotted against $E(B-V)$, with sources using integrated dust maps marked with triangles (as lower limits on $(B-V)_0$). At peak, we see a negative slope of $-0.2$ in the best fit line to this bronze sample. The p-value for this case is 0.07, just above the typical cut-off of 0.05, and not quite low enough to make this negative slope statistically significant (compared against a null hypothesis of zero slope). At $t_2$, we see another negative slope when fitting the bronze sample, with a p-value of 0.008. If we restrict this analysis to the silver sample, then the p-values become larger, with some evidence for a non-zero slope in the case of the colours at $t_2$. We note, however, that this correlation is driven largely by one nova here with the highest reddening, which perhaps has a slightly underestimated reddening. The correlation disappears if this point is not included, or if our reddening is underestimated by $1 \sigma$. As a result, we do not consider this correlation to be a significant effect. It is most likely the case that the slope detected with the full sample is a result of only having limits on some of the colours for novae using integrated dust map reddenings.

\subsubsection{Uncertainty Analysis}

\begin{figure}
 \includegraphics[width=\columnwidth]{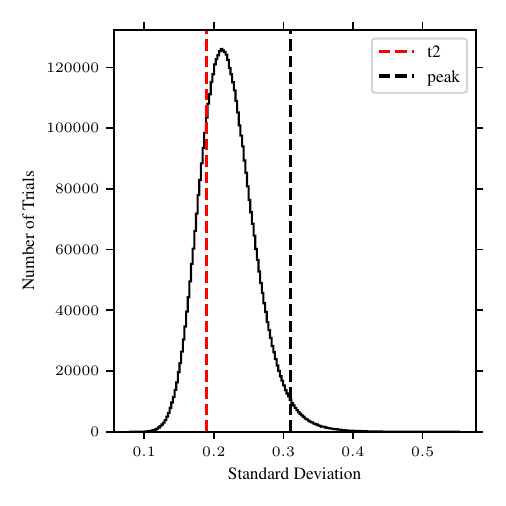}
 \caption{ Monte Carlo simulation results displaying the expected standard deviation of $(B-V)_0$ for our silver sample. This is the result of five million random samples across our photometric and reddening uncertainties. The dashed red line indicates the standard deviation of our silver sample at $t_2$, and the dashed black line indicates the result at $V$-band peak. This particular simulation used the sample of novae at peak, although a very similar result is obtained from the sample at $t_2$.}
 \label{fig:mcsim}
\end{figure}

With these distributions in hand, next we want to know if the scatter in the intrinsic colour is dominated by the uncertainties in our reddenings and photometry, or if it contains a meaningful contribution from intrinsic colour variations between novae. To test this, we have run a Monte Carlo simulation with five million samples to sample over our uncertainties. For each sample, we loop through the novae in the silver sample, and randomly draw a colour based on the estimated reddening and photometric uncertainty, combined with the mean colour. For each iteration, we compute the standard deviation of the sample, and then examine how frequently we expect to get values similar to those that we observe. The resulting distribution of the standard deviations may be seen in Figure \ref{fig:mcsim}. Most of this scatter is driven by the reddening uncertainties, so future data sets with better constrained estimates of $E(B-V)$ will allow for better constraints on the intrinsic nova variability. The samples at peak and $t_2$ are slightly different due to different availability of photometry. As a result, we've performed this analysis for each case and find similar statistics between the two.

All of these samples are done for the $(B-V)_0$ colour. Both $(R-I)_0$ and $(V-R)_0$ are anticipated to have similar results, with similar overall uncertainties. In each of these cases, the reddening will be a bit lower, but we gain additional uncertainty from the assumed reddening law used in measuring the colour excess for these colours. The results for the silver sample at peak show that a standard deviation of $0.23$ is expected given our uncertainties. For the sample at $t_2$, this predicts $0.22$ instead, fully explaining the scatter at $t_2$. Our scatter at peak is not so easily explained, and likely includes some intrinsic variability. The scatter in the reddenings should be the same between the two cases, so it is unlikely to be responsible for the larger scatter at peak. The larger peak scatter is also recovered in the gold sample, where the reddenings are the same at peak and $t_2$. Similarly, the photometric accuracy is generally expected to be superior at peak, as the novae are two magnitudes brighter. This simulation indicates that the probability of achieving a 0.3 standard deviation or larger at peak is $6 \%$. It is therefore indicated that there is intrinsic variability in the nova colours around peak, but not at $t_2$.

\subsection{Do Nova Colours Correlate with Other Properties?}\label{sec:correlation}

Nova colours at peak and $t_2$ show a scatter of $\pm 0.2-0.3$ mag (Table \ref{tab:colourstats}); do nova colour differences correlate with other properties of the eruptions?

\begin{figure}
 \includegraphics[width=\columnwidth]{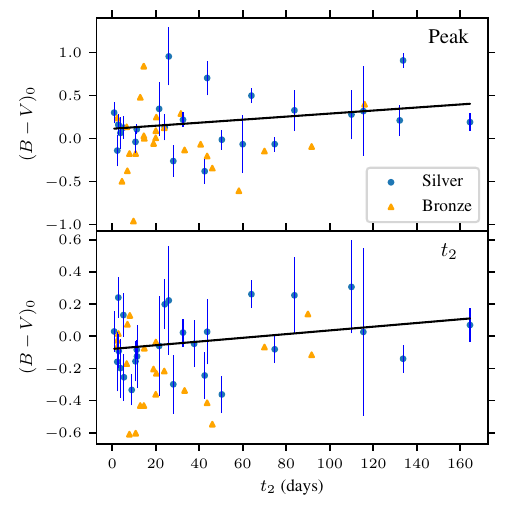}
 \caption{$(B-V)_0$ plotted against $t_2$ for the novae in our sample. The top panel displays the colour at the optical peak. and the bottom panel uses the colour at $t_2$. The black lines display a line of best fit made to the silver sample, with a near zero slope in both panels. Blue points represent novae in the silver sample with $E(B-V)$ from DIBs or 3D dust maps, while orange triangles use integrated dust maps.  All sources combined yield the bronze sample. The fit at peak has a p-value of 0.17 and a Pearson correlation coefficient of 0.23. At $t_2$, the p-value for the silver sample is 0.16, with a Pearson correlation coefficient of 0.28.}
 \label{fig:bv_vs_t2}
\end{figure}

Figure~\ref{fig:bv_vs_t2} shows the correlations of $(B-V)_0$ plotted against the speed class of the nova, $t_2$. There are no significant correlations between $(B-V)_0$ and $t_2$, either for our DIB or dust map samples, or at peak or $t_2$. The p-value is $> 0.15$ at both peak and $t_2$ for the silver sample, indicating that there is no significant evidence in our sample for non-zero slopes. \cite{Hachisu2016} suggest that nova colours may be correlated with the nova speed class, specifically that faster novae ought to be bluer. This also implies that the colour scales with the ejecta mass of the eruption. Our data set does not provide a significant detection of such a correlation.

\begin{figure}
 \includegraphics[width=\columnwidth]{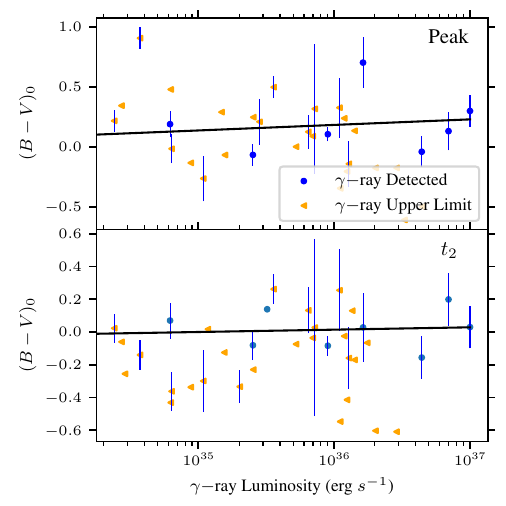}
 \caption{ The top panel displays the peak $(B-V)_0$ plotted against the nova's GeV $\gamma$-ray luminosity (or upper limit on luminosity). The black line displays a line of best fit made to the novae in the silver sample that have $\gamma$-ray detections. The bottom panel is similar, but displays the colours at $t_2$. Orange triangles in this figure displays sources that have upper limits on the $\gamma$-ray luminosity, but are not $\gamma$-ray detected. Novae belonging to the silver sample have y-axis uncertainties, while those sources without uncertainties plotted use 2D dust maps. The fit at peak has a p-value of 0.76 and a Pearson correlation coefficient of 0.14. At $t_2$, the p-value for the silver sample is 0.72, with a Pearson correlation coefficient of 0.14.}
 \label{fig:bv_vs_gamma}
\end{figure}

\begin{figure}
 \includegraphics[width=\columnwidth]{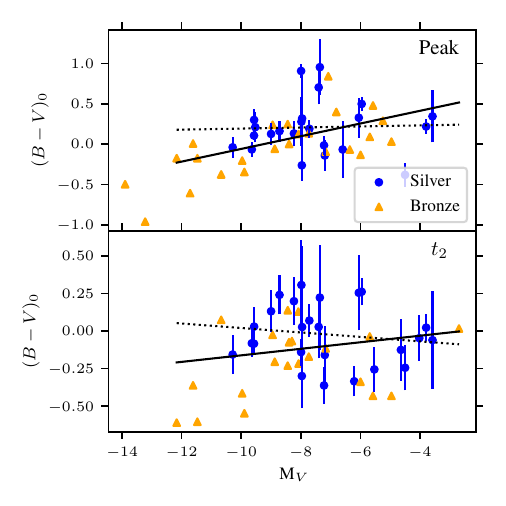}
 \caption{ The top panel displays the peak $(B-V)_0$ plotted against the estimated $V$-band absolute magnitude at peak. The solid black line displays a linear best fit line, and the dotted line shows a fit made only to the silver sample. Here the blue points belong to the silver sample, while the orange triangles are limits on the colour derived using the 2D dust map reddenings. The bottom panel is similar, but displays the colours at $t_2$. The fit at peak to the silver sample has a p-value of 0.87 and a Pearson correlation coefficient of 0.04. At $t_2$, the p-value for the silver sample is 0.49, with a Pearson correlation coefficient of -0.15.}
 \label{fig:bv_vs_MV}
\end{figure}

We have also checked for correlations of colour with the nova luminosity in the optical and GeV $\gamma$-ray bands. 
The shocks responsible for accelerating $\gamma$-ray emitting particles
\citep{2018A&A...612A..38M,2022A&A...660A.104D} may
contribute to the nova optical emission \citep{2017NatAs...1..697L,2017MNRAS.469.4341M,2020NatAs...4..776A} in principle, affecting its colour.
Figure~\ref{fig:bv_vs_gamma} plots $(B-V)_0$ against the average $>$100 MeV $\gamma$-ray luminosity, measured (or constrained to an upper limit) by the Large Area Telescope on the \emph{Fermi Gamma-Ray Space Telescope}. We use the average $\gamma$-ray fluxes (or upper limits) collected by Craig et al.\ 2025 (in preparation), which makes heavy use of the analyses of \citet{Ackermann2014}, \citet{Cheung2016}, and \citet{Franckowiak2018}, amongst others. There is not evidence for a significant correlation in either case, with p-values of 0.76 and 0.72 at peak and $t_2$, respectively.

The $V$-band absolute magnitude at peak is plotted against $(B-V)_0$ in Figure~\ref{fig:bv_vs_MV}. The black line shows a best-fitting line, which has a significant non-zero slope (p-value of 0.0004) for the colours measured at peak. However, the non-zero slope is driven by the dust map sources, and is likely not be real, as novae in the lower left of this figure may suffer from overestimated $E(B-V)$ from the 2D dust maps. This will tend to decrease both the colour and the absolute magnitude for these novae at both peak and $t_2$. A fit made with only novae that have DIB-based $E(B-V)$ measurements is shown with the dotted black line, where the slope is much closer to zero. Without the dust map novae, the non-zero slope is no longer significant.

In the above figures, we only search for correlations with $(B-V)_0$, but a similar analysis is carried out in the Appendix for $(V-R)_0$ and $(R-I)_0$. No significant correlations are found for these colour indices when excluding 2D dust map limits, similar to our above findings for $(B-V)_0$. With the bronze sample included, both colours have significant correlations with the reddening, as a result of the integrated map usage.

\subsection{Colour Evolution} \label{sec:colourev}

For the novae in our sample, we monitor the evolution of their colours to search for trends. Based on spectroscopy of nova eruptions, emission lines become increasingly important flux sources in the optical as the nova evolves. Around the peak, the bulk of the optical flux is expected to be continuum emission from the nova photosphere, with relatively minimal contributions from emission lines. Sample spectra at peak and $t_2$ with the $BVRI$ filter passbands is shown in Figure \ref{fig:samplespec}. This continuum flux should be fairly similar between novae, and explains the similar colours observed at around peak. As the emission line flux increases, we ought to see more variability in the colours depending on emission line strengths.

\begin{figure*}
 \includegraphics[]{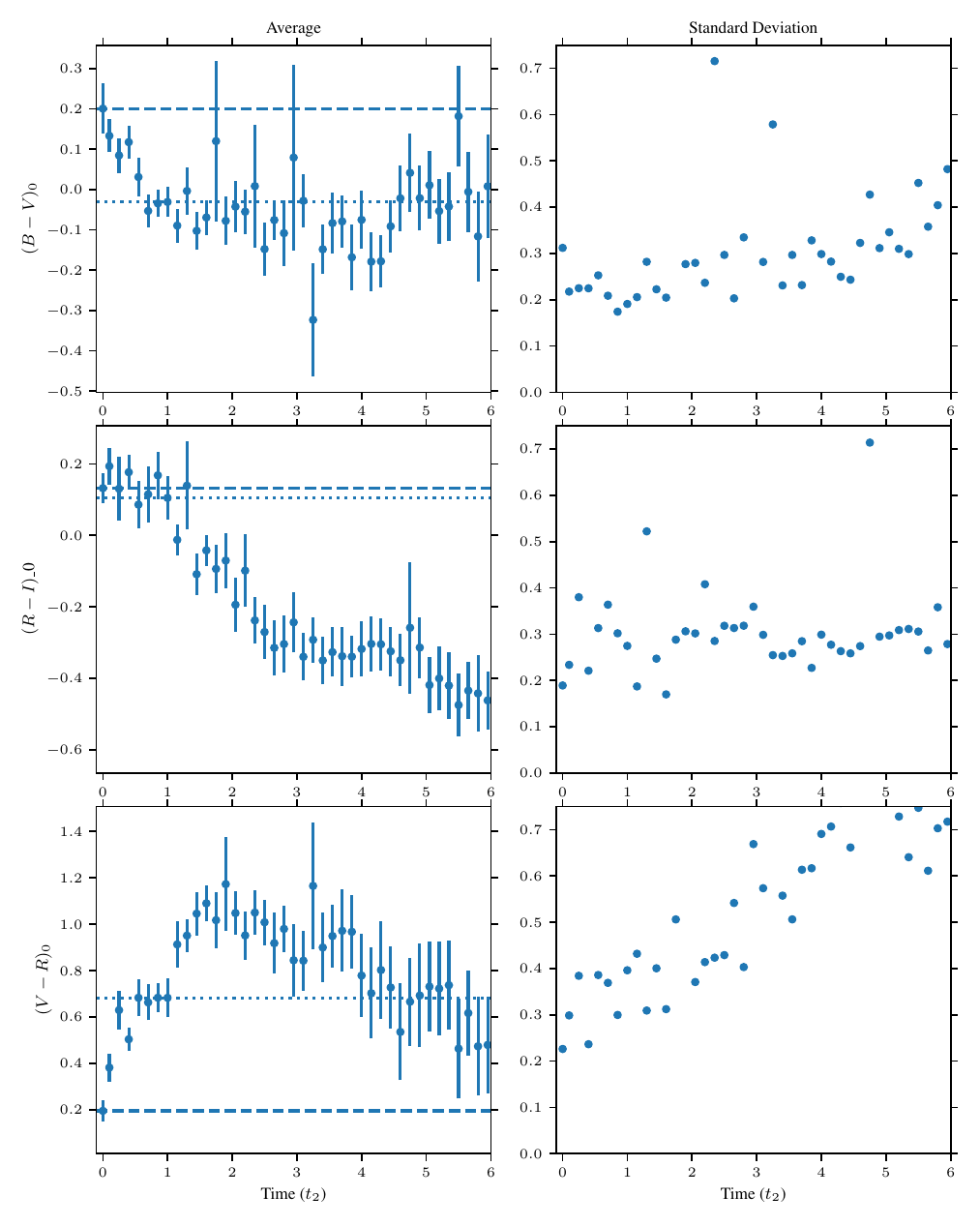}
 \caption{Evolution of the nova colours during eruption, plotted in units of $t_2$. The left column shows the average colour (with error bars set as the standard error on the mean) for the three colours that we monitor. The right column displays the standard deviations for the sample at each time. In each panel in the left column, the dashed line shows the colour at peak, and the dotted line shows the colour at $t_2$ for comparison at later parts of the light curve.}
 \label{fig:colour_evolution}
\end{figure*}

Figure~\ref{fig:colour_evolution} displays the average colour for the silver sample, with the standard error on the mean, and the standard deviation of the colours for our nova sample. In $(B-V)_0$ we see that the colour begins at 0.20 at peak ($t = 0$), then becomes bluer approaching $t_2$. At late times, we see some variability in the colour, but it typically remains close to $(B-V)_0 = 0$. At $t > 2t_2$, the standard deviation in $(B-V)_0$ begins to rise, consistent with our expectations from emission lines contributions in the spectra.

In $(R-I)_0$, the colour is reddest around the peak, similar to $(B-V)_0$. Over time, the colour gets bluer, perhaps with a small plateau at around $t_2$ with a colour close to 0. Beyond $2 t_2$, the colour starts getting bluer again, with a small rise in the dispersion in the colours. After $3 t_2$, $(R-I)_0$ flattens out to a constant at around -0.34. $(R-I)_0$ shows the most stable standard deviation of the three colour indices examined here. $(V-R)_0$ shows large amounts of early time evolution, with $(V-R)_0$ becoming significantly redder after the peak. It then plateaus at just past $t_2$, until it trends bluer again at late times. The colour beyond three times $t_2$ has a large amount of variability between novae, possibly caused by the large emission line contributions in some bands.

For the purposes of estimating reddening with $(B-V)$ photometry, a variety of times around $t_2$ are suitable with average colours consistently close to 0. This even holds approaching several times $t_2$, and could potentially be used for novae detected after the peak. It then does become challenging to suitably estimate the uncertainties, since the scatter in the colours grows at later times, and without an identification of the peak it is difficult to assess where along this curve a measurement sits. Nonetheless, this may remain useful as a method for measuring reddening towards fading novae that lack another estimate. If the nova is still within $\sim 3$ times $t_2$ of the peak, then constraints on the reddening with uncertainties of $0.3$ magnitudes or less are possible. Later time colours for $(R-I)_0$ and $(V-R)_0$ show more evolution compared to $(B-V)_0$, and should only be used if it is clear when the nova peak was; otherwise the evolution in these colours could lead to misleading reddenings.

\subsection{Using Nova Colours to Infer Distances}\label{sec:distance}

To systematically constrain the distances to the novae in our sample, we compare the measured foreground reddening with three-dimensional dust maps, in this case using the 3D maps of \cite{Chen2019}. In addition, we used the Galactic mass model described in \citet[][see also \citealt{Kawash2021b}]{Kawash2022} to further constrain the distances. We generated a population of $10^7$ simulated novae in this model, distributed in proportion to stellar mass as described by a realistic  three-dimensional model of the Milky Way \citep{Robin2003}. Then for each simulated nova, we calculated its foreground $V$-band extinction using a composite three-dimensional dust model as implemented in {\tt mwdust} \citep{Bovy2016}, which combines several three-dimensional dust maps \citep{Green2019, Marshall2006, Drimmel2003}. As in \citet{Kawash2022}, we assume that the simulated novae have peak absolute magnitudes normally distributed with mean $M_V = -7.2, \sigma=0.8$ mag, but truncated at higher luminosities for longer $t_2$ values. The simulated absolute magnitude is combined with the distance and line-of-sight extinction to infer simulated peak $V$ magnitudes.

Here, we do not use the simulated apparent magnitudes in our distance inference, but instead just use the typical distances to novae in a given direction. This constraint is primarily useful for cases where the nova is beyond all of the Galactic extinction in the 3D maps, and therefore the dust maps provide only a limit on the distance. It would also be possible to further constrain the distance if we used the assumed optical luminosity function with the measured peak apparent mag and E(B-V). The observed magnitudes of novae can then be compared with that of nearby novae in the simulations to provide a distance estimate that utilises the nova luminosity function. We choose not to do this here, as distances are often wanted to constrain the luminosities of novae, and therefore distances derived under assumptions about luminosity are in danger of circular reasoning. Take, for example, our comparison of optical and $\gamma$-ray properties in Craig et al. 2025, in prep. It is desirable to test if $\gamma$-ray luminous novae are otherwise normal in their optical luminosity, and estimating optical/gamma-ray luminosities based on a distance, which is in turn based on assumptions about luminosity, would invalidate this comparison. For this reason, we only use extinction to estimate distances here.

Our measurements of the average colours of novae at peak and $t_2$ allow for estimates of $E(B-V)$ from photometry alone, with typical accuracies of $\sim 0.2$ mag (translating to $\sim$0.6 mag uncertainty on dust extinction, $A_V$).
When combined with 3D dust maps, this can provide a distance estimate for the nova, making use of these photometrically derived extinctions. The accuracy of these inferred distances will depend on how the extinction evolves along the line of sight towards the nova, and may only offer a lower limit on the distance if the nova is beyond the bulk of the Galactic reddening. We have estimated the distances towards the novae in our sample based on our best available extinction estimates. These utilise DIB measurements where possible, and use our photometric colour calibrations where spectra with sufficient resolution are not available.

\begin{figure}
 \includegraphics[width=\columnwidth]{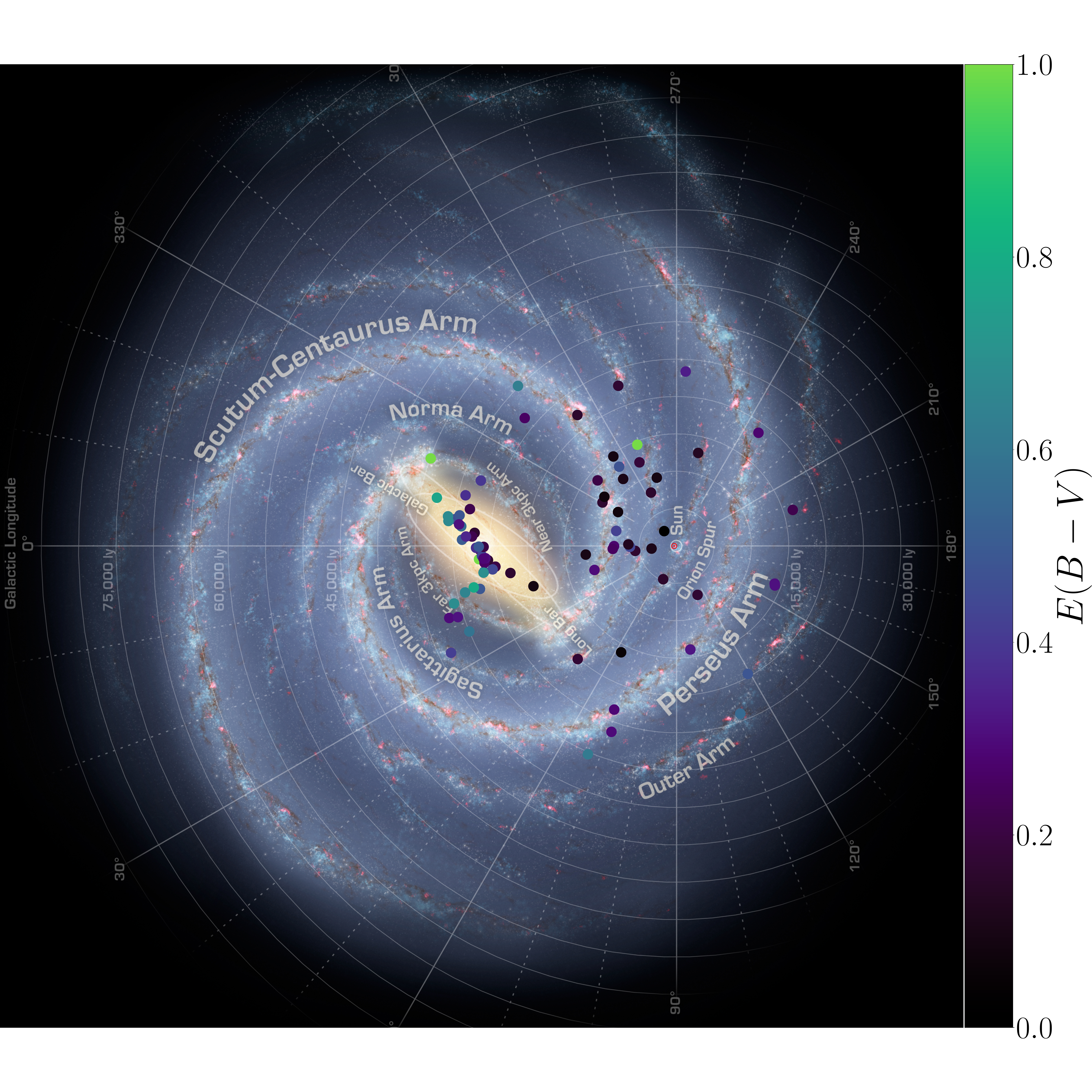}
 \caption{Locations of our nova sample plotted over an artist's image of the MW (R. Hurt: NASA/JPL-Caltech/SSC). The novae are colour coded based on their $E(B-V)$ values, and have distances derived using their measured $E(B-V)$ combined with 3D dust maps, coupled with their expected distance based on the distribution of stars in the Galaxy.}
 \label{fig:MW}
\end{figure}

The distances derived here are included in Table \ref{tab:allnovae}, along with distances from \emph{Gaia} DR3 where parallax measurements are $> 3 \sigma$. Distances used for calculating absolute magnitudes or luminosities will be the parallax distance when it is available, otherwise we use the distances determined here. A plot of these novae over a top-down artist's image of the Milky Way is shown in Figure~\ref{fig:MW}. There are many novae in the bulge, as is expected, especially since novae with distances driven by a MW mass model in that direction will tend to be placed in the bulge. In the outer parts of the disc, we see the novae following the spiral arms, which is not likely to be a real effect, but is due to the large extinctions in the spiral arms. For a distance determined using extinctions, most extinction values will correspond to distances in the spiral arms, so systematically distances in the arms are preferred.

\begin{figure}
 \includegraphics[width=\columnwidth]{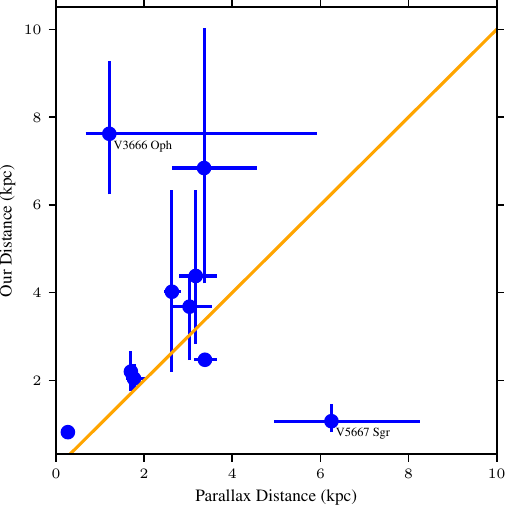}
 \caption{Distances measured using our extinction values and 3D dust maps, plotted against \emph{Gaia} parallax distance measurements. All of the parallaxes are for novae with $> 3 \sigma$ parallax measurements. The solid line is a unity line where the distances are equivalent.}
 \label{fig:Distance_Comp}
\end{figure}

To check the accuracy of these extinction-derived distances with \emph{Gaia} parallaxes, we plot our distance estimates against the \emph{Gaia} DR3 parallax distances in Figure~\ref{fig:Distance_Comp}. Only sources where the parallax divided by the parallax uncertainty is greater than 3 are included ($\varpi / \sigma_{\varpi} > 3$), similar to the cut used in \citep{Schaefer2022}. The parallax distances are derived using the methods of \citep{Bailer-Jones2018}, with the length scales set following \citep{Schaefer2018}. We can see that the distances agree within the uncertainties for most sources, with a few exceptions. 

The novae with the largest discrepancies are V3666 Oph and V5667 Sgr. Our distance to V5667 Sgr is driven by a DIB-based extinction measurement, from which 3D dust maps suggest a small distance of 1.4 kpc. The \emph{Gaia} parallax instead implies a much larger distance of 6.8 kpc, and is inconsistent with a distance of 1.4 kpc. In this case, we prefer the parallax measurement for use in our calculations, to avoid problems relating to inaccuracies in the 3D dust map for this particular source (possibly due to small scale variations in the extinction in this direction). For V3666 Oph, our extinction estimate suggests a distance larger than 3.2 kpc, while the \emph{Gaia} parallax provides a distance of 1.2 kpc. The extinction towards V3666 Oph is estimated as larger than the maximum extinction from the 3D dust map, thereby providing a limit on the distance. A distance of 3.2 kpc would be consistent with the \emph{Gaia} parallax within 1 $\sigma$.

\begin{figure}
 \includegraphics[width=\columnwidth]{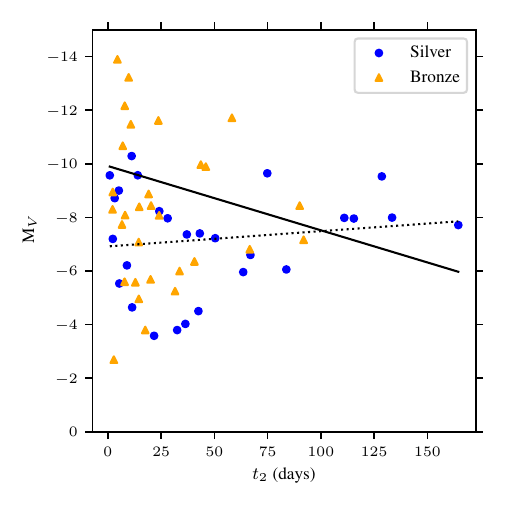}
 \caption{MMRD for our sample of Galactic novae, plotting the absolute magnitude at peak against $t_2$. There is a significant amount of scatter in this figure, with a small (and not statistically significant) negative slope based on linear fitting. The blue points use DIB based $E(B-V)$ measurements, while the orange triangles are from integrated dust maps. The orange triangles are therefore effectively lower limits on the absolute magnitude. The solid line is fit through the full nova sample, while the dotted line is fit to the silver sample only. The fit to the silver sample has a p-value of 0.50 and a Pearson correlation coefficient of -0.14}.
 \label{fig:MMRD}
\end{figure}

With our sample of novae that include measurements for $t_2$ and $V$-band absolute magnitude at peak ($M_V$), we can plot the MMRD relation for our sample. This is shown in Figure \ref{fig:MMRD}. A large amount of scatter is clear in this relation, and the correlation is not significant in our data set, with a p-value of 0.43 for the bronze sample (and 0.50 with the silver sample only). Large scatter is typical for the MMRD, and is evident even with larger samples, as found by many papers (e.g. \citealt{Buscombe1955,Arp1956,Shafter2011,Ozdonmez2018,Schaefer2022,Clark2024}). We have also checked the distribution of the absolute magnitudes at peak, and see a mean of $-7.8$ with a standard deviation of 2 magnitudes. The mean is similar to other results \citep{Shafter2017}, with a larger spread compared to other work. It is likely the case that the increased spread is due to a combination of limited sample size, and larger relative uncertainties in the distances.

\section{Conclusions} \label{sec:concl}

We gathered $BVRI$ photometry for 61 recent Galactic novae observed by AAVSO and SMARTS \citep{Walter2012}, in order to systematically survey the optical colours of recent Galactic novae (\S \ref{sec:sample}).

Corrections for interstellar reddening are accomplished using $E(B-V)$ measurements made using methods entirely independent of the nova photometry (presented in Table \ref{tab:allnovae}). We prefer using DIB absorption to measure the interstellar reddening in the nova's spectrum (\S \ref{sec:dib}). We find that $E(B-V)$ estimated from \ion{Na}{i} D absorption has a large dispersion (compared to DIB-based values) in our medium-resolution ($R \approx 9000$) spectra, and should be treated with caution (\S \ref{sec:naid}). In cases where \emph{Gaia} parallax is measured to $> 3 \sigma$ significance, we use 3D dust maps to estimate $E(B-V)$, and in other cases, 2D dust maps provide useful upper limits on $E(B-V)$ (\S \ref{sec:dustmaps}).

By utilizing the large set of publicly available nova photometry, we can measure nova colours across a variety of bands, with a focus here on optical $BVRI$ photometry. As demonstrated in \S \ref{sec:colourdist}, novae in our sample exhibit $(B-V)_0$ colours at their $V$ band peaks that typically vary by $\pm$0.3 magnitudes, and exhibit somewhat smaller spreads in their colours closer to $t_2$. Similar spreads in $(V-R)_0$ and $(R-I)_0$ colours are observed (as described by the  statistics in Table \ref{tab:colourstats}). These measured distributions of nova colours may be utilised for estimating reddening towards other novae, especially those that lack high quality spectra.
Colour-based reddening measurements only depend on availability of light curves, and those can be fairly limited in cadence. Such data are often available for novae, and can provide a useful estimate of the extinction towards the nova, with A$_{V}$ precise to $\sim$0.6\,mag. Based on the work in this paper, reddening can be estimated using $(B-V)$, $(R-I)$ or $(V-R)$, and we recommend using the values from the silver samples as given in Table \ref{tab:colourstats}, which have the best combination of reliable $E(B-V)$ measurements and sample size. There is not a significant difference in colour spread between these colour indices, and since $B$ and $V$ light curves tend to be better sampled, $(B-V)$ will likely be the more useful colour. \textbf{Our recommendation is to use $\mathbf{(B-V)_0 = -0.2 \pm 0.3}$ at peak and $\mathbf{(B-V)_0 = 0.0\pm 0.2}$ at $\mathbf{t_2}$ for estimating the reddening towards novae.} At peak, our data indicates that there is some intrinsic variability in $(B-V)_0$, while at $t_2$ the observed scatter can be readily explained by the reddening and measurement uncertainties. As the nova eruption evolves, emission line contributions to the colours are expected to become significant and begin to increase the colour diversity (\S \ref{sec:colourev}; Figure \ref{fig:colour_evolution}). $(V-R)_0$ especially shows significant evolution in the average colour, and the standard deviation increases by approximately a factor of 3 over time.

We combine our $E(B-V)$ measurements with 3D dust maps to estimate distances to the novae in our sample, using a novel Bayesian technique that takes the mass distribution of the Milky Way as a prior (\S \ref{sec:distance}; Table \ref{tab:allnovae}). Figure~\ref{fig:MW} shows a face-on map of the Milky Way with our nova sample superimposed, demonstrating that novae prefer the Galactic bulge, and we generally miss novae on the far side of the Galaxy.

Nova colours across our sample do not show any statistically significant correlations with $E(B-V)$, $t_2$, the peak $V$-band absolute magnitude, or $\gamma$-ray luminosity (\S \ref{sec:correlation}). The lack of colour correlation with $t_2$ implies that the nova's ejecta mass, which generally influences $t_2$, has minimal impact on the observed colours. Nova colours, especially $(B-V)_0$, tend to be relatively consistent between nova eruptions, unlike many other nova properties. These calibrations can be further refined in the future, when a larger sample of novae with suitable spectra for making DIB measurements becomes available.

\section*{Acknowledgements}

We are grateful for support from NASA grants \#80NSSC23K0497 and \#80NSSC23K1247, along with NSF grants AST-2107070 and AST-2205628. E.A. acknowledges support by NASA through the NASA Hubble Fellowship grant HST-HF2-51501.001-A awarded by the Space Telescope Science Institute, which is operated by the Association of Universities for Research in Astronomy, Inc., for NASA, under contract NAS5-26555. 

We acknowledge with thanks the variable star observations from the AAVSO International Database contributed by observers worldwide and used in this research. We thank the following ARAS observers, who took spectra that were utilised in this work; Peter Somogyi, Terry Bohlsen, Ernst Pollmann, Keith Graham, Olivier Thizy, Thierry Lemoult, Fran\c{c}ois Teyssier, E.~Bryssinck, Etienne Bertrand, Colin Eldridge, Tim Lester, Benjamin Mauclaire, WR13x-collaboration, Umberto Sollecchia, Jack Martin, Jacques Montier, J. Powles, Michel Verlinden, Robin Leadbeater, Jean-Noel Terry, Jose Ribeiro, D. Grennan, Paul Luckas, Dave Doctor, Terry Bohlsen, Mariusz Bajer, T. Hansen, Marion Thomas, JGFFMT, Sean Curry, E. Barbotin, Christian Kreider, Dong Li, Paolo Berardi, and A. Favaro.

We especially thank Christian Buil, Berto Monard, Thibault de France, Margaret Streamer and Jeremy Shears for contributing large amounts of data useful for this work.

Our analysis made use of \textsc{IRAF}, \textsc{Python3}, \textsc{numpy},
\textsc{scipy}, \textsc{matplotlib}, \textsc{mw plot}, and \textsc{mwdust} This work has made use of colour maps from the \textsc{CMasher} package \cite{cmasher}.

\section*{Data Availability}
All photometry used in this work is publicly available through the AAVSO website or the SMARTS atlas for (mostly) southern novae. The majority of the spectra used for $E(B-V)$ measurements are available through SMARTS or ARAS, with the exception of a small number of SALT spectra. The measurements from the SALT spectra required to reproduce our statistics are provided in Table \ref{tab:allnovae}, in the form of $E(B-V)$ measurements. Our colour measurements in all of our colours are also provided in Table \ref{tab:allnovae}. All of the colour and $E(B-V)$ measurements used here are available in Table \ref{tab:allnovae}.



\bibliographystyle{mnras_vanHack}
\bibliography{nova_bib} 



\clearpage
\appendix

\section{Additional Data and Figures}\label{app}

\begin{figure}
 \includegraphics[width=\columnwidth]{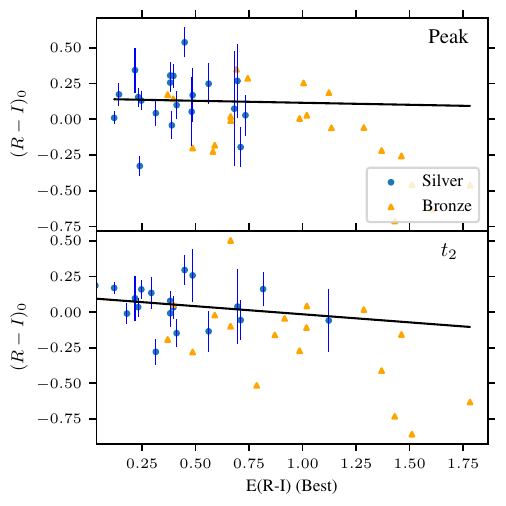}
 \caption{ The top panel displays the peak $(R-I)_0$ plotted against the estimated $E(R-I)$ values. The black line displays a linear best fit to the silver sample. The bottom panel is similar, but displays the colours at $t_2$. Orange triangles in these figures show measurements made using the integrated dust map, and the blue points with uncertainties shown are from the silver sample. The fit at peak has a p-value of 0.91 and a Pearson correlation coefficient of -0.03. At $t_2$, the p-value for the silver sample is 0.19, with a Pearson correlation coefficient of -0.31.}
 \label{fig:ri_vs_ebv}
\end{figure}

\begin{figure}
 \includegraphics[width=\columnwidth]{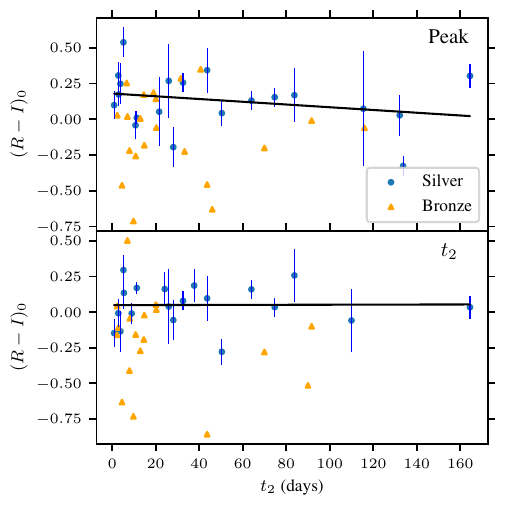}
 \caption{$(R-I)_0$ plotted against $t_2$ for the novae in our sample. The top panel displays the colour at the optical peak, and the bottom panel uses the colour at $t_2$ instead. The black lines display a line of best fit, with a near zero slope in both panels with the fit made to the silver sample. Blue points are from the silver sample, and orange triangles are limits on the colour from integrated dust maps. The fit at peak has a p-value of 0.29 and a Pearson correlation coefficient of -0.25. At $t_2$, the p-value for the silver sample is 0.45, with a Pearson correlation coefficient of -0.18.}
 \label{fig:ri_vs_t2}
\end{figure}

\begin{figure}
 \includegraphics[width=\columnwidth]{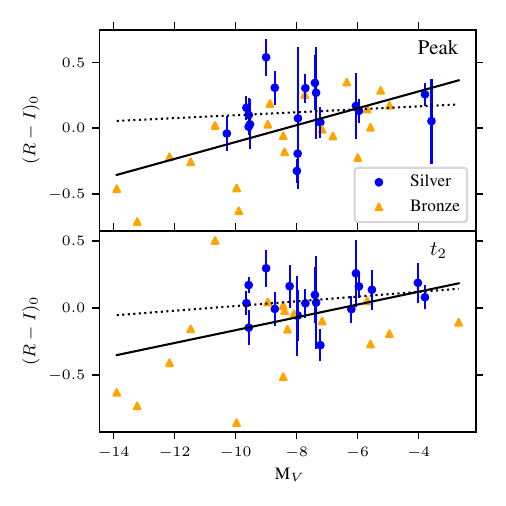}
 \caption{ The top panel displays the peak $(R-I)_0$ plotted against the estimated $V$-band absolute magnitude at peak. The solid black line displays a linear best fit line, and the dotted line shows a fit made only to the silver sample. Here, the blue dots represent novae from the silver sample, while the orange triangles show novae with integrated dust map reddenings. The bottom panel is similar, but displays the colours at $t_2$. The fit at peak to the silver sample has a p-value of 0.68 and a Pearson correlation coefficient of 0.11. At $t_2$, the p-value for the silver sample is 0.38, with a Pearson correlation coefficient of 0.22.}
 \label{fig:ri_vs_MV}
\end{figure}

\begin{figure}
 \includegraphics[width=\columnwidth]{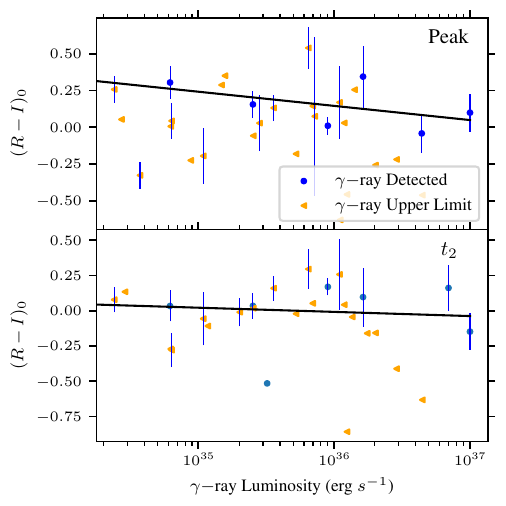}
 \caption{ The top panel displays the peak $(R-I)_0$ plotted against the $\gamma$-ray luminosity (or upper limit on the luminosity). The black line displays a linear best fit to the novae in the silver sample that are $\gamma$-ray detected. The bottom panel is similar, but displays the colours at $t_2$. Orange triangles in these figures show novae with limits on the $\gamma$-ray luminosity, while blue points show $\gamma$-ray detections. Points with uncertainties belong to the silver sample, while points without uncertainties utilise integrated dust map reddenings. The fit at peak has a p-value of 0.31 and a Pearson correlation coefficient of -0.50. At $t_2$, the p-value for the silver sample is 0.73, with a Pearson correlation coefficient of -0.15}.
 \label{fig:ri_vs_gamma}
\end{figure}

The correlations of the $(R-I)_0$ measurements against various nova parameters are shown here in Figures~\ref{fig:ri_vs_ebv}, \ref{fig:ri_vs_t2}, \ref{fig:ri_vs_MV}, and \ref{fig:ri_vs_gamma}. None of these cases show a statistically significant correlation, unless the bronze sample is used, in which case the correlations are considered to be unreliable due to the inclusion of 2D dust maps reddenings. The strongest signals appear with the absolute magnitudes, however this may be related to the distance estimates towards these novae that depend on the estimated extinctions, such that the two measurements are not independent.

\begin{figure}
 \includegraphics[width=\columnwidth]{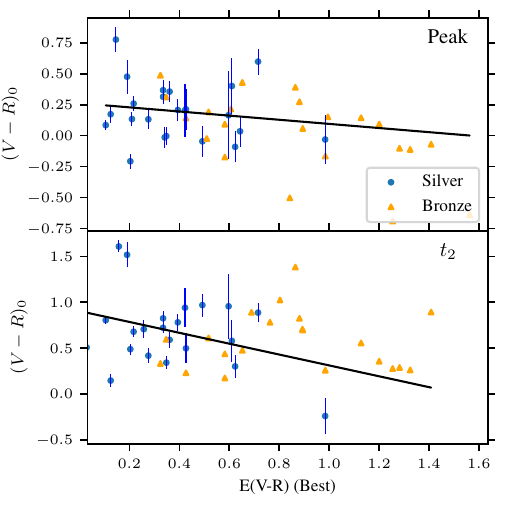}
 \caption{ The top panel displays the peak $(V-R)_0$ plotted against the estimated $E(V-R)$ values. The black line displays a linear best fit, made to the novae in the silver sample. The bottom panel is similar, but displays the colours at $t_2$. Orange triangles in these figures show measurements made using integrated dust maps, and the blue points with uncertainties shown are based on spectral measurements or 3D dust maps. The fit at peak has a p-value of 0.47 and a Pearson correlation coefficient of -0.16. At $t_2$, the p-value for the silver sample is 0.13, with a Pearson correlation coefficient of -0.33.}
 \label{fig:vr_vs_ebv}
\end{figure}

\begin{figure}
 \includegraphics[width=\columnwidth]{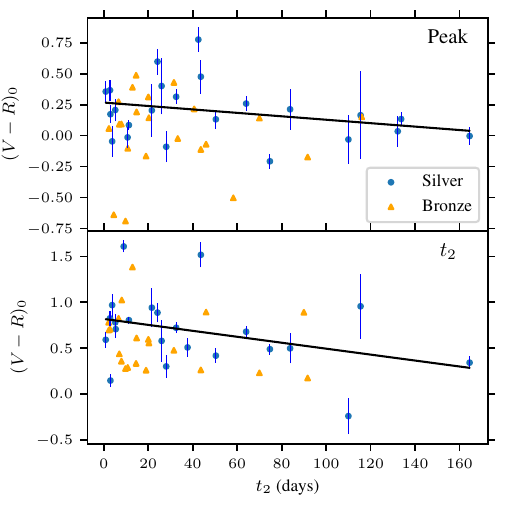}
 \caption{$(V-R)_0$ plotted against $t_2$ for the novae in our sample. The top panel displays the colour at the optical peak. and the bottom panel uses the colour at $t_2$ instead. The black line displays a linear best fit, made to the novae in the silver sample. Blue points have DIB or 3D dust map measurements of $E(B-V)$, while orange triangles use integrated dust maps. The fit at peak has a p-value of 0.18 and a Pearson correlation coefficient of -0.29. At $t_2$, the p-value for the silver sample is 0.10, with a Pearson correlation coefficient of -0.35.}
 \label{fig:vr_vs_t2}
\end{figure}

\begin{figure}
 \includegraphics[width=\columnwidth]{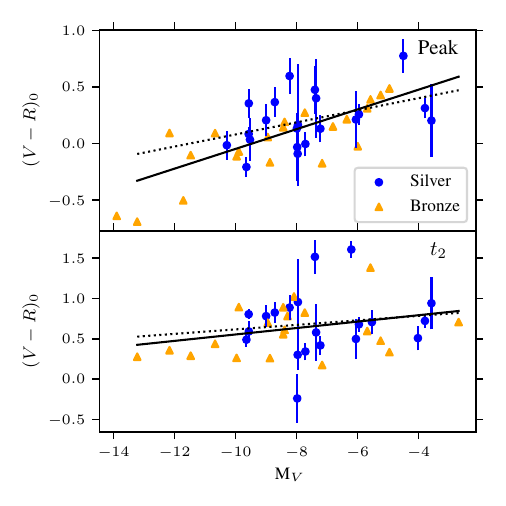}
 \caption{ The top panel displays the peak $(V-R)_0$ plotted against the estimated $V$-band absolute magnitude at peak. The solid black line displays a linear best fit line, and the dotted line shows a fit made only to the silver sample. Here, the blue dots represent novae from the silver sample, while the orange triangles show novae with integrated dust map reddenings. The bottom panel is similar, but displays the colours at $t_2$. The fit at peak to the silver sample has a p-value of 0.051 and a Pearson correlation coefficient of 0.43. At $t_2$, the p-value for the silver sample is 0.58, with a Pearson correlation coefficient of -0.0.13.}
 \label{fig:vr_vs_MV}
\end{figure}

\begin{figure}
 \includegraphics[width=\columnwidth]{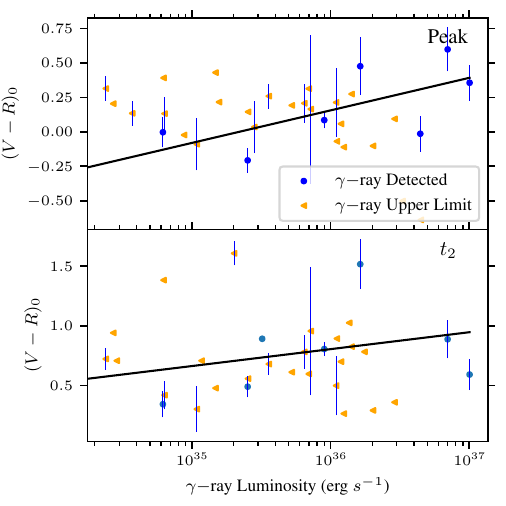}
 \caption{The top panel displays the peak $(V-R)_0$ plotted against the $\gamma$-ray luminosity (or upper limit on the luminosity). The black line displays a linear best fit to the novae in the silver sample that are $\gamma$-ray detected. The bottom panel is similar, but displays the colours at $t_2$. Orange triangles in these figures show novae with limits on the $\gamma$-ray luminosity, while blue points show $\gamma$-ray detections. Points with uncertainties belong to the silver sample, while points without uncertainties utilise integrated dust map reddenings. The fit at peak has a p-value of 0.12 and a Pearson correlation coefficient of 0.64. At $t_2$, the p-value for the silver sample is 0.27, with a Pearson correlation coefficient of 0.45.}
 \label{fig:vr_vs_gamma}
\end{figure}

The correlations of the $(V-R)_0$ measurements against various nova parameters are shown here in Figures~\ref{fig:vr_vs_ebv}, \ref{fig:vr_vs_t2}, \ref{fig:vr_vs_MV}, and \ref{fig:vr_vs_gamma}. The results here are quite similar to those in $(B-V)_0$ and $(R-I)_0$.

Table \ref{tab:allnovae} contains the $E(B-V)$ values for all of the novae in our sample, and provides $(B-V)_0$, $(R-I)_0$, and $(V-R)_0$ intrinsic colours both at peak and $t_2$ for each nova. The listed $E(B-V)$ is the best value produced by our spectroscopic measurements, 3D dust maps, and 2D dust maps. We also include our distance measurements, estimated from $E(B-V)$ as described in \S \ref{sec:distance}, and \emph{Gaia} parallax distances where $> 3 \sigma$ measurements are available.

\renewcommand{\arraystretch}{1.25}
\begin{table*}
\caption{Details of our sample, including the name of each nova, the time of the optical peak, $t_2$, the $E(B-V)$ estimates used in this work, and our colour measurements corrected for interstellar reddening. Uncertainties on the optical peak time are derived based on the cadence of the data around the peak. In the rare case that the resulting uncertainty is less than  0.01 days, we round up to 0.01.} The uncertainty on $E(B-V)$ is provided when available, mainly for cases with spectroscopic ``DIB" measurements. All of the dust map cases do not have uncertainties listed. Note that there are two novae here, V392 Per and V1535 Sco, which are not part of our colour samples due to an unconstrained peak time. They are included here however because they have both parallax distances and reddening based distances, and are included in the comparison between these distance measurements.\label{tab:allnovae}
\setlength{\tabcolsep}{3pt}
\begin{tabular}{|c|c|c|c|c|c|c|c|c|c|c|c|c|}

\hline
Name & Optical Peak & $t_2$ & $E(B-V)$ & Method & $(B-V)_{0}$ & $(B-V)_{0}$ & $(R-I)_{0}$ & $(R-I)_{0}$ & $(V-R)_{0}$ & $(V-R)_{0}$ & Distance & Parallax Dist. \\
     & (UT) & (days) & & & peak & $t_2$ & peak &$t_2$ & peak & $t_2$& (kpc) & (kpc)\\
\hline
U Sco & 2022-06-$7.1_{-0.17}^{+0.11}$ & 3.0 & 0.19 $\pm$ 0.11 & DIB & 0.15 & $-0.10$ & 0.17 & 1.14 & 0.17 & 0.15 & $6.6^{+-10.6}_{-3.4}$ & -- \\
 RS Oph & 2021-08-$10.1_{-0.01}^{+0.05}$ & 3.8 & 0.75 $\pm$ 0.19 & DIB & 0.04 & $-0.22$ & 0.25 & $-0.14$ & $-0.05$ & 0.97 & $6.0^{+-7.5}_{-2.6}$ & $2.6^{+0.2}_{-0.1}$ \\ 
 V606 Vul & 2021-08-$2.1_{-0.02}^{+0.13}$ & 91.7 & 0.89 & 2D Map & $-0.12$ & $-0.15$ & $-0.01$ & $-0.10$ & $-0.17$ & 0.18 & $7.3^{+-9.7}_{-3.6}$ & -- \\ 
 V1711 Sco & 2021-06-$25.2_{-1.92}^{+3.02}$ & 23.9 & 1.39 & 2D Map & 0.07 & $-0.27$ & -- & -- & -- & -- & $9.3^{+-11.4}_{-6.8}$ & -- \\ 
 V1674 Her & 2021-06-$13.0_{-0.01}^{+0.14}$ & 0.9 & 0.55 $\pm$ 0.13 & DIB & 0.28 & 0.01 & 0.10 & $-0.15$ & 0.36 & 0.59 & $6.3^{+-4.3}_{-2.7}$ & -- \\ 
 V1710 Sco & 2021-04-$13.6_{-0.21}^{+0.53}$ & 4.5 & 2.38 & 2D Map & $-0.58$ & $-1.21$ & $-0.46$ & $-0.63$ & $-0.64$ & -- & $10.2^{+-13.0}_{-5.7}$ & -- \\ 
 V6595 Sgr & 2021-04-$6.2_{-0.50}^{+0.43}$ & 5.2 & 0.60 $\pm$ 0.14 & DIB & 0.10 & 0.11 & 0.54 & 0.30 & 0.21 & 0.78 & $8.2^{+-7.1}_{-6.2}$ & -- \\ 
 V1405 Cas & 2021-05-$10.3_{-0.17}^{+0.19}$ & 164.5 & 0.53 $\pm$ 0.11 & DIB & 0.17 & 0.05 & 0.30 & 0.03 & 0.00 & 0.34 & $2.2^{+-1.8}_{-1.7}$ & $1.7^{+0.1}_{-0.1}$ \\ 
 V1112 Per & 2020-11-$28.2_{-0.01}^{+0.08}$ & 28.1 & 0.95 $\pm$ 0.19 & DIB & $-0.30$ & $-0.33$ & $-0.19$ & $-0.06$ & $-0.09$ & 0.30 & $4.3^{+-2.3}_{-2.5}$ & -- \\ 
 V6593 Sgr & 2020-10-$4.8_{-0.06}^{+0.03}$ & 19.0 & 1.50 & 2D Map & $-0.11$ & $-0.26$ & 0.19 & -- & $-0.16$ & 0.26 & $8.1^{+-9.7}_{-4.8}$ & -- \\ 
 V1391 Cas & 2020-08-$10.9_{-0.08}^{+0.06}$ & 110.0 & 1.5 $\pm$ 0.30 & DIB & 0.23 & 0.25 & -- & $-0.06$ & $-0.03$ & $-0.24$ & $6.1^{+-3.6}_{-2.4}$ & -- \\ 
 V6567 Sgr & 2020-06-$16.4_{-0.69}^{+0.36}$ & 116.0 & 1.52 & 2D Map & 0.35 & -- & $-0.06$ & -- & 0.15 & -- & $8.4^{+-10.7}_{-4.4}$ & -- \\ 
 V659 Sct & 2019-10-$31.1_{-0.33}^{+0.66}$ & 7.9 & 1.83 & 2D Map & $-0.24$ & $-0.67$ & $-0.22$ & $-0.41$ & 0.09 & 0.36 & $9.3^{+-12.0}_{-3.8}$ & -- \\ 
 V2891 Cyg & 2019-11-$29.0_{-1.0}^{+3.96}$ & 70.0 & 3.91 & 2D Map & -- & -- & -- & -- & $-1.21$ & $-1.08$ & -- & -- \\ 
 V3890 Sgr & 2019-08-$28.1_{-0.01}^{+0.01}$ & 2.8 & 0.51 $\pm$ 0.13 & DIB & 0.14 & 0.22 & 0.31 & $-0.01$ & 0.37 & 0.83 & $7.1^{+-5.8}_{-5.3}$ & -- \\ 
 V3666 Oph & 2018-08-$11.9_{-0.1}^{+0.88}$ & 21.6 & 0.64 $\pm$ 0.32& 3D Map & 0.32 & $-0.08$ & 0.05 & -- & 0.20 & 0.94 & $7.6^{+-9.0}_{-6.0}$ & $1.2^{+4.7}_{-0.5}$ \\ 
 V613 Sct & 2018-07-$1.0_{-0.01}^{+0.3}$  & 26.0 & 0.93 $\pm$ 0.35 & DIB & 0.92 & 0.19 & 0.27 & 0.04 & 0.40 & 0.58 & $9.9^{+-6.9}_{-5.1}$ & -- \\ 
 V392 Per & 2018-04-$29.8_{-8.34}^{+0.12}$ & 3.0 & 0.96 $\pm$ 0.30 & DIB & -- & -- & -- & -- & -- & -- & $4.4^{+-2.8}_{-2.4}$ & $3.2^{+0.5}_{-0.4}$ \\ 
 V5857 Sgr & 2018-04-$10.7_{-0.95}^{+0.75}$ & 20.0 & 2.44 & 2D Map & $-0.08$ & $-0.45$ & -- & -- & -- & -- & $8.6^{+-11.1}_{-3.2}$ & -- \\ 
 V435 CMa & 2018-03-$29.0_{-0.53}^{+0.42}$ & 60.0 & 0.87 $\pm$ 0.35 & DIB & $-0.10$ & -- & -- & -- & -- & -- & $5.8^{+-4.2}_{-2.9}$ & -- \\ 
 V906 Car & 2018-03-$28.5_{-0.92}^{+0.99}$ & 43.7 & 0.29 $\pm$ 0.21 & DIB & 0.69 & 0.02 & 0.34 & 0.10 & 0.48 & 1.52 & $3.0^{+-2.2}_{-2.0}$ & -- \\ 
 V1662 Sco & 2018-02-$9.8_{-0.51}^{+0.56}$ & 11.5 & 20.52 & 2D Map & $-18.29$ & $-18.83$ & $-14.14$ & $-13.84$ & $-11.94$ & $-11.39$ & $10.9^{+-14.1}_{-5.8}$ & -- \\ 
 FM Cir & 2018-01-$28.3_{-0.95}^{+0.97}$ & 133.7 & 0.32 $\pm$ 0.09 & DIB & 0.90 & $-0.15$ & $-0.33$ & -- & 0.13 & -- & $3.6^{+-2.4}_{-2.9}$ & -- \\ 
 V549 Vel & 2017-10-$17.7_{-2.35}^{+3.98}$ & 90.0 & 1.05 & 2D Map & -- & 0.10 & -- & $-0.52$ & -- & 0.89 & $7.3^{+-10.4}_{-3.7}$ & -- \\ 
 V612 Sct & 2017-07-$30.0_{-1.02}^{+0.86}$ & 132.2 & 0.98 $\pm$ 0.19 & DIB & 0.18 & -- & 0.03 & -- & 0.04 & -- & $9.6^{+-6.2}_{-5.2}$ & -- \\ 
 V5856 Sgr & 2016-11-$8.0_{-0.21}^{+0.39}$ & 10.7 & 0.52 $\pm$ 0.13 & DIB & $-0.06$ & $-0.17$ & $-0.04$ & -- & $-0.01$ & -- & $7.9^{+-6.5}_{-6.5}$ & -- \\ 
 V407 Lup & 2016-09-$24.4_{-0.42}^{+0.99}$ & 9.0 & 0.24 $\pm$ 0.10 & DIB & -- & $-0.34$ & -- & $-0.01$ & -- & 1.61 & $2.9^{+-2.1}_{-2.1}$ & -- \\ 
 V5669 Sgr & 2015-09-$28.5_{-0.45}^{+0.95}$ & 33.3 & 0.78 & 2D Map & $-0.16$ & $-0.37$ & $-0.23$ & -- & $-0.02$ & -- & $2.7^{+-2.3}_{-1.9}$ & -- \\ 
 V2944 Oph & 2015-04-$13.6_{-1.00}^{+1.04}$ & 32.5 & 0.51 $\pm$ 0.09 & DIB & 0.20 & 0.01 & 0.26 & 0.08 & 0.31 & 0.72 & $1.8^{+-1.4}_{-1.5}$ & -- \\ 
 V5668 Sgr & 2015-03-$21.4_{-0.26}^{+0.26}$ & 74.7 & 0.31 $\pm$ 0.09 & DIB & $-0.08$ & $-0.09$ & 0.16 & 0.04 & $-0.21$ & 0.49 & $3.9^{+-3.2}_{-3.5}$ & -- \\ 
 V5667 Sgr & 2015-02-$15.7_{-0.95}^{+2.01}$ & 64.0 & 0.33 $\pm$ 0.09 & DIB & 0.49 & 0.25 & 0.13 & 0.16 & 0.26 & 0.68 & $1.1^{+-0.8}_{-0.7}$ & $6.2^{+2.0}_{-1.3}$ \\ 
 V1535 Sco & 2015-02-11.0 & 15.0 & 0.69 $\pm$ 0.19 & DIB & -- & -- & -- & -- & -- & -- & $8.8^{+-6.8}_{-6.4}$ & $6.2^{+3.4}_{-0.7}$ \\ 
 V2659 Cyg & 2014-04-$10.5_{-0.74}^{+0.48}$ & 115.5 & 0.91 $\pm$ 0.54 & DIB & 0.29 & $-0.01$ & 0.07 & -- & 0.16 & 0.96 & $8.2^{+-4.6}_{-4.7}$ & -- \\ 
 V962 Cep & 2014-03-$13.9_{-0.79}^{+0.48}$ & 31.5 & 0.99 & 2D Map & 0.26 & -- & 0.29 & -- & 0.43 & 0.48 & $4.4^{+-6.1}_{-2.1}$ & -- \\ 
 V5666 Sgr &  2014-02-$17.4_{-0.99}^{+1.0}$ & 70.0 & 0.65 & 2D Map & $-0.17$ & $-0.09$ & $-0.20$ & $-0.28$ & 0.14 & 0.23 & -- & -- \\ 
 V1369 Cen & 2013-12-$14.7_{-1.93}^{+2.03}$ & 37.7 & 0.04 $\pm$ 0.15 & 3D Map & -- & $-0.05$ & -- & 0.19 & -- & 0.51 & $0.8^{+-0.9}_{-0.8}$ & $0.3^{+0.1}_{-0.1}$ \\ 
 V339 Del & 2013-08-$16.7_{-0.46}^{+0.09}$ & 11.3 & 0.16 $\pm$ 0.06 & DIB & 0.10 & $-0.09$ & 0.01 & 0.17 & 0.09 & 0.81 & $5.1^{+-3.4}_{-2.6}$ & -- \\ 
 V809 Cep & 2013-02-$4.7_{-0.63}^{+0.07}$  & 20.3 & 1.72 & 2D Map & 0.19 & $-0.29$ & $-0.06$ & 0.02 & 0.15 & 0.56 & $7.4^{+-10.4}_{-3.5}$ & -- \\ 
 V5593 Sgr & 2012-07-$22.4_{-0.18}^{+7.06}$ & 43.6 & 2.02 & 2D Map & $-0.28$ & $-0.49$ & $-0.46$ & $-0.86$ & $-0.11$ & 0.26 & $9.0^{+-11.7}_{-4.3}$ & -- \\ 
 V5591 Sgr & 2012-06-$27.3_{-0.67}^{+0.18}$ & 2.3 & 1.36 & 2D Map & 0.19 & $-0.07$ & 0.03 & 0.04 & 0.06 & 0.70 & $7.7^{+-9.1}_{-4.9}$ & -- \\ 
 V1324 Sco & 2012-06-$20.0_{-0.28}^{+0.36}$ & 24.2 & 1.09 $\pm$ 0.16 & DIB & 0.09 & 0.16 & -- & 0.16 & 0.60 & 0.89 & $8.8^{+-10.2}_{-7.1}$ & -- \\ 
 V2677 Oph & 2012-05-$21.3_{-2.6}^{+0.17}$ & 6.7 & 1.34 & 2D Map & 0.09 & $-0.22$ & 0.26 & -- & 0.27 & 0.82 & $8.1^{+-9.5}_{-6.9}$ & -- \\ 

 \hline
\end{tabular}
\end{table*}

\begin{table*}
\contcaption{Details of our sample, including the name of each nova, the time of the optical peak, $t_2$, the $E(B-V)$ estimates used in this work, and our colour measurements corrected for interstellar reddening. Uncertainties on the optical peak time are derived based on the cadence of the data around the peak. In the rare case that the resulting uncertainty is less than  0.01 days, we round up to 0.01. The uncertainty on $E(B-V)$ is provided when available, mainly for cases with spectroscopic ``DIB" measurements. All of the dust map cases do not have uncertainties listed. Note that there are two novae here, V392 Per and V1535 Sco, which are not part of our colour samples due to an unconstrained peak time. They are included here however because they have both parallax distances and reddening based distances, and are included in the comparison between these distance measurements.}
\setlength{\tabcolsep}{3pt}
\begin{tabular}{|c|c|c|c|c|c|c|c|c|c|c|c|c|}

\hline
Name & Optical Peak & $t_2$ & $E(B-V)$ & Method & $(B-V)_{0}$ & $(B-V)_{0}$ & $(R-I)_{0}$ & $(R-I)_{0}$ & $(V-R)_{0}$ & $(V-R)_{0}$ & Distance & Parallax Dist. \\
     & (UT) & (days) & & & peak & $t_2$ & peak &$t_2$ & peak & $t_2$& (kpc) & (kpc)\\
\hline
 V5589 Sgr & 2012-04-$22.5_{-1.61}^{+0.14}$ & 2.3 & 0.8 $\pm$ 0.19 & DIB & $-0.17$ & $-0.19$ & -- & -- & -- & -- & $8.0^{+-9.4}_{-6.3}$ & -- \\ 
 V1428 Cen & 2012-04-$7.8_{-0.43}^{+0.46}$ & 10.8 & 1.95 & 2D Map & $-0.24$ & $-0.67$ & $-0.26$ & $-0.16$ & $-0.10$ & 0.29 & $9.4^{+-12.6}_{-5.2}$ & -- \\ 
 V2676 Oph & 2012-04-$4.5_{-1.04}^{+1.89}$ & 83.8 & 0.65 $\pm$ 0.25 & DIB & 0.30 & 0.23 & 0.17 & 0.26 & 0.21 & 0.50 & $8.1^{+-9.7}_{-6.5}$ & -- \\ 
 V834 Car & 2012-03-$1.4_{-0.21}^{+0.10}$ & 20.1 & 0.53 & 2D Map & 0.07 & $-0.06$ & 0.14 & 0.05 & 0.31 & 0.60 & $7.1^{+-9.8}_{-3.7}$ & -- \\ 
 V1313 Sco & 2011-09-$7.5_{-0.38}^{+0.51}$ & 8.1 & 1.22 & 2D Map & -- & 0.09 & -- & $-0.04$ & -- & 1.02 & $8.6^{+-11.3}_{-4.1}$ & -- \\ 
 PR Lup & 2011-08-$14.4_{-0.24}^{+0.67}$ & 14.7 & 0.79 & 2D Map & $-0.03$ & $-0.10$ & $-0.18$ & $-0.02$ & 0.19 & 0.61 & $8.3^{+-11.4}_{-4.0}$ & -- \\ 
 V1312 Sco & 2011-06-$2.3_{-3.33}^{+0.18}$ & 12.9 & 1.32 & 2D Map & 0.43 & $-0.48$ & 0.01 & $-0.27$ & 0.39 & 1.38 & $2.6^{+-3.1}_{-2.4}$ & -- \\ 
 T Pyx & 2011-05-$12.0_{-0.16}^{+0.31}$ & 50.4 & 0.42 $\pm$ 0.12 & DIB & $-0.03$ & $-0.38$ & 0.04 & $-0.28$ & 0.13 & 0.42 & $4.0^{+-2.2}_{-1.7}$ & $2.6^{+0.2}_{-0.2}$ \\ 
 V5588 Sgr & 2011-04-$7.1_{-0.98}^{+1.02}$ & 46.0 & 2.15 & 2D Map & $-0.42$ & $-0.62$ & $-0.63$ & $-1.25$ & $-0.07$ & 0.89 & $8.1^{+-9.9}_{-3.6}$ & -- \\ 
 V1311 Sco & 2010-04-$26.3_{-0.57}^{+0.82}$ & 2.2 & 1.16 & 2D Map & -- & $-0.11$ & -- & $-0.16$ & -- & 0.78 & $9.1^{+-11.6}_{-6.1}$ & -- \\ 
 V496 Sct & 2009-11-$18.5_{-1.01}^{+0.99}$ & 58.2 & 1.28 & 2D Map & $-0.65$ & -- & -- & -- & $-0.50$ & -- & $10.4^{+-14.4}_{-6.1}$ & -- \\ 
V5584 Sgr & 2009-11-$3.4_{-2.01}^{+1.4}$ & 40.7 & 0.92 & 2D Map & $-0.10$ & -- & 0.35 & -- & 0.22 & -- & $3.6^{+-4.6}_{-2.9}$ & -- \\ 
V2672 Oph & 2009-08-$16.5_{-2.37}^{+0.06}$ & 2.8 & 1.36 & 2D Map & -- & $-0.03$ & -- & $-0.11$ & -- & 0.71 & $2.0^{+-2.1}_{-1.9}$ & -- \\ 
 V5583 Sgr & 2009-08-$6.4_{-0.08}^{+0.09}$ & 5.4 & 0.39 $\pm$ 0.15 & 3D Map & -- & $-0.27$ & -- & 0.13 & -- & 0.71 & $2.0^{+-2.3}_{-1.7}$ & $1.8^{+0.2}_{-0.2}$ \\ 
 V1213 Cen & 2009-05-$8.2_{-4.07}^{+2.87}$ & 11.4 & 0.50 $\pm$ 0.21 & 3D Map & -- & $-0.14$ & -- & -- & -- & -- & $6.8^{+-9.5}_{-3.6}$ & $3.4^{+1.2}_{-0.7}$ \\ 
 V679 Car & 2008-11-$29.7_{-4.37}^{+0.06}$ & 16.2 & 6.10 & 2D Map & $-5.50$ & $-5.52$ & -- & -- & -- & -- & $4.5^{+-5.0}_{-4.0}$ & -- \\ 
 V2670 Oph & 2008-05-$27.4_{-0.92}^{+0.57}$ & 14.5 & 0.94 & 2D Map & 0.81 & -- & -- & -- & -- & -- & $8.0^{+-9.1}_{-6.8}$ & -- \\ 
 V5579 Sgr & 2008-04-$23.4_{-0.61}^{+0.14}$ & 7.0 & 0.89 & 2D Map & $-0.41$ & 0.04 & 0.02 & 0.50 & 0.09 & 0.44 & $8.1^{+-9.3}_{-5.9}$ & -- \\ 
 V2491 Cyg & 2008-04-$11.4_{-2.55}^{+0.06}$ & 14.5 & 0.49 & 2D Map & 0.01 & $-0.45$ & 0.17 & $-0.19$ & 0.49 & 0.33 & $1.5^{+-2.0}_{-1.2}$ & -- \\ 
 NR TrA & 2008-04-$13.7_{-0.93}^{+0.87}$  & 42.5 & 0.22 $\pm$ 0.15 & 3D Map & $-0.39$ & $-0.25$ & -- & -- & 0.78 & -- & $3.7^{+-4.9}_{-3.0}$ & $3.0^{+0.5}_{-0.4}$ \\ 
 V2468 Cyg & 2008-03-$9.5_{-1.02}^{+0.7}$ & 9.8 & 1.91 & 2D Map & -1.03 & -1.38 & -0.71 & -0.73 & -0.69 & 0.28 & $9.4^{+-12.8}_{-6.2}$ & -- \\ 
  \hline
\end{tabular}
\end{table*}

\bsp     
\label{lastpage}

\end{document}